\documentclass[10pt,twocolumn]{article}
\usepackage{cvpr}
\usepackage{times}
\usepackage{epsfig}
\usepackage{graphicx}
\usepackage{amsmath}
\usepackage{amssymb}

 \cvprfinalcopy % *** Uncomment this line for the final submission
\def\cvprPaperID{9174} % *** Enter the CVPR Paper ID here

\ifcvprfinal\fi

\begin{document}

%%%%%%%%% TITLE
\title{Pathological Retinal Region Segmentation From OCT Images Using Geometric Relation Based Augmentation}

% \author{Dwarikanath Mahapatra\\
% Inception Institute of Artificial Intelligence\\
% Abu Dhabi, United Arab Emirates\\
% {\tt\small dwarikanath.mahapatra@inceptioniai.org}
% % For a paper whose authors are all at the same institution,
% % omit the following lines up until the closing ``}''.
% % Additional authors and addresses can be added with ``\and'',
% % just like the second author.
% % To save space, use either the email address or home page, not both
% \and
% Behzad Bozorgtabar \\
% Ecole polytechnique fédérale de Lausanne\\
% Switzerland\\
% {\tt\small behzad.bozorgtabar@epfl.ch}
% \and
% Ling Shao\\
% Inception Institute of Artificial Intelligence\\
% Abu Dhabi, United Arab Emirates\\
% {\tt\small ling.shao@inceptioniai.org}
% }

\author{Dwarikanath Mahapatra$^{1}$ \\
\and
Behzad Bozorgtabar$^{2,3,4}$ \\
\and
Jean-Philippe Thiran$^{2,3,4}$ \\ 
\and
Ling Shao$^{1}$\\
\and
$^{1}$ IIAI, Abu Dhabi \quad \quad 
$^{2}$ LTS5, EPFL, Lausanne \quad \quad
$^{3}$ CIBM, Laussane \quad \quad
$^{4}$ CHUV, Lausanne \\
%$^{1}$ Inception Institute of Artificial Intelligence  (IIAI), Abu Dhabi, UAE \\
%$^{2}$\'Ecole Polytechnique F\'ed\'erale de Lausanne (EPFL), Laussane, Switzerland \\
%$^{3}$\ Centre d'Imagerie Biom\'edicale (CIBM), Laussane, Switzerland \\
%$^{4}$\ Department of Radiology, Centre Hospitalier
%Universitaire Vaudois (CHUV), Lausanne, Switzerland \\
{\tt\small \{firstname.lastname\}@inceptioniai.org,@epfl.ch}
%{\tt\small dwarikanath.mahapatra@inceptioniai.org, behzad.bozorgtabar@epfl.ch, ling.shao@inceptioniai.org}
}

\maketitle
%\thispagestyle{empty}

%%%%%%%%% ABSTRACT
\begin{abstract}
   Medical image segmentation is an important task for computer aided diagnosis. Pixelwise manual annotations of large datasets require high expertise and is time consuming. Conventional data augmentations have limited benefit by not fully representing the underlying distribution of the training set, thus affecting model robustness when tested on images captured from different sources. Prior work leverages synthetic images for data augmentation ignoring the interleaved geometric relationship between different anatomical labels. We propose improvements over previous GAN-based medical image synthesis methods by jointly encoding the intrinsic relationship of geometry and shape. Latent space variable sampling results in diverse generated images from a base image and improves robustness. Given those augmented images generated by our method, we train the segmentation network to enhance the segmentation performance of retinal optical coherence tomography (OCT) images. The proposed method outperforms state-of-the-art segmentation methods on the public RETOUCH dataset having images captured from different acquisition procedures. Ablation studies and visual analysis also demonstrate benefits of integrating geometry and diversity.

   %Augmented datasets using our method for automatic segmentation of retinal optical coherence tomography (OCT) images outperform existing methods on the public RETOUCH dataset having images captured from different acquisition procedures.  
\end{abstract}

%%%%%%%%% BODY TEXT

\section{Introduction}
\label{sec:intro}

\begin{figure}[t]
 \centering
\begin{tabular}{ccc}
\includegraphics[height=2.2cm, width=2.4cm]{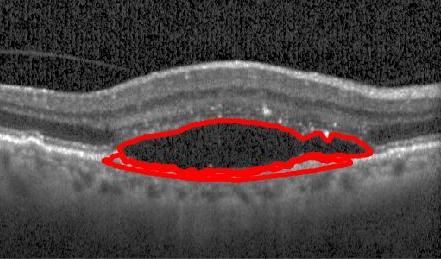} &
\includegraphics[height=2.2cm, width=2.4cm]{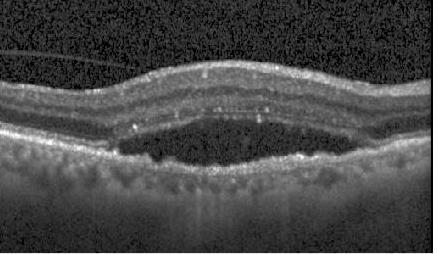} &
\includegraphics[height=2.2cm, width=2.4cm]{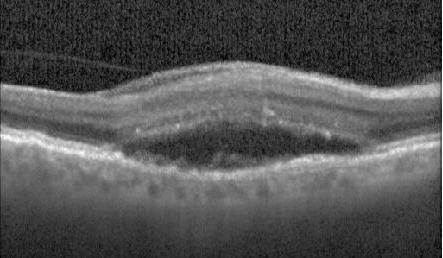} \\
(a) & (b) & (c) \\
\end{tabular}
\begin{tabular}{cc}
\includegraphics[height=2.6cm, width=3.4cm]{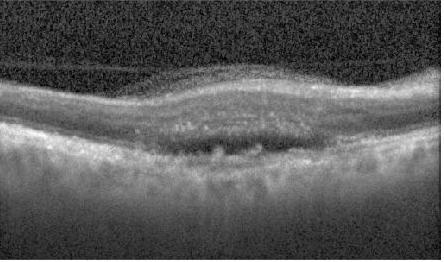} &
\includegraphics[height=2.6cm, width=3.4cm]{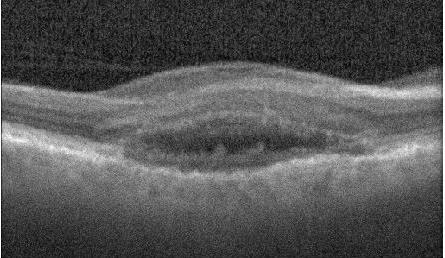} \\
 (d) & (e) \\
\end{tabular}
\caption{(a) Base image (red contour denotes segmentation mask); Example of generated images using: (b) Our proposed $GeoGAN$ method; (c) Zhao et al. \cite{Zhao_CVPR2019}; (d) $DAGAN$ method by \cite{DAGAN}; (e) $cGAN$ method by \cite{Mahapatra_MICCAI2018}. }
\label{fig:SynImages_comp}
\end{figure}

Medical image segmentation is an important task for healthcare applications like disease diagnosis, surgical planning, and disease progression monitoring. While deep learning (DL) methods demonstrate state-of-the-art results for medical image analysis tasks \cite{DLRevTMI,Mahapatra_PR2020,ZGe_MTA2019,Behzad_PR2020,Mahapatra_CVIU2019,Mahapatra_CMIG2019}, their robustness depends upon the availability of a diverse training dataset to learn different disease attributes such as appearance and shape characteristics. 
 Large scale dataset annotations for segmentation require image pixel labels, which is time consuming and involves high degree of clinical expertise. The problem is particularly acute for pathological images since it is difficult to obtain diverse images for less prevalent disease conditions, necessitating data augmentation. We propose a generative adversarial network (GAN) based approach for pathological images augmentation and demonstrate its efficacy in pathological region segmentation. Figure~\ref{fig:SynImages_comp} summarizes the image generation results of our approach and \cite{Zhao_CVPR2019,Mahapatra_LME_PR2017,Zilly_CMIG_2016,Mahapatra_SSLAL_CD_CMPB,Mahapatra_SSLAL_Pro_JMI,Mahapatra_LME_CVIU,LiTMI_2015}, and highlights our superior performance by incorporating geometric information.

\begin{figure}[ht]
\begin{center}
   \begin{tabular}{c}
\includegraphics[height=3.8cm, width=7.2cm]{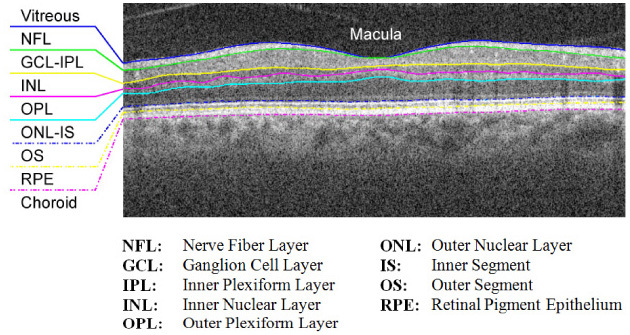} \\
(a) \\
\includegraphics[height=2.6cm, width=7.2cm]{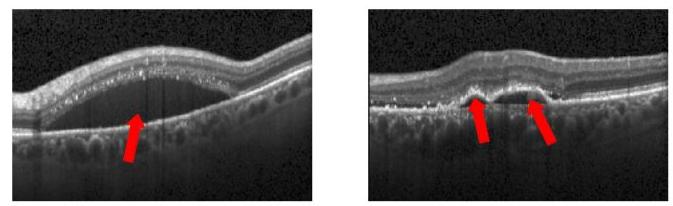}  \\
(b) \\
\end{tabular}
\end{center}
   \caption{Example of normal and fluid filled OCT images: (a) example control subject image without any abnormalities (taken from \cite{OCTFig}); (b) images with accumulated fluid up due to diabetic macular edema and AMD from our dataset. The fluid areas are marked with red arrows.}
\label{fig:OCT1}
\end{figure}

Traditional augmentations such as image rotations or deformations have limited benefit as they do not fully represent the underlying data distribution of the training set and are sensitive to parameter choices. Recent works \cite{huang2018auggan,Zhao_CVPR2019, han2018gan, nielsen2019gan,MahapatraJDI_Cardiac_FSL,Mahapatra_JSTSP2014,MahapatraTIP_RF2014,MahapatraTBME_Pro2014,MahapatraTMI_CD2013,MahapatraJDICD2013} proposed to solve this issue by using synthetic data for augmentation and increase diversity in the training samples. However, certain challenges have not been satisfactorily addressed by these methods. 

 Zhao et. al. \cite{Zhao_CVPR2019,MahapatraJDIMutCont2013,MahapatraJDIGCSP2013,MahapatraJDIJSGR2013,MahapatraJDISkull2012,MahapatraTIP2012,MahapatraTBME2011,MahapatraEURASIP2010} proposed a learning-based registration method to register images to an atlas, use  corresponding deformation field to deform a segmentation mask and obtain new data. This approach presents the following challenges: 1) since registration errors propagate to subsequent stages, inaccurate registration can adversely affect the data generation process; 2) with atlas of a normal subject it is challenging to register images from diseased subjects due to appearance or shape changes. This is particularly relevant for layer segmentation in retinal optical coherence tomography (OCT) images, where there is a drastic difference in layer shape between normal and diseased cases. Figure~\ref{fig:OCT1} (a) shows the retinal layers of a normal subject, and Figure~\ref{fig:OCT1} (b) shows two cases of retinal fluid build up due to diabetic macular edema (DME) and age related macular degeneration (AMD). The retinal layers are severely distorted compared to Figure~\ref{fig:OCT1} (a) and registration approaches have limited impact in generating accurate images. 

Recent methods for data augmentation \cite{han2018gan, nielsen2019gan, Mahapatra_CVIU19, bozorgtabar2019learn,Kuanar_ICIP19,Bozorgtabar_ICCV19,Xing_MICCAI19,Mahapatra_ISBI19,MahapatraAL_MICCAI18,Mahapatra_MLMI18,Sedai_OMIA18} using a generative adversarial network (GAN) \cite{goodfellow2014generative,Sedai_MLMI18,MahapatraGAN_ISBI18,Sedai_MICCAI17,Mahapatra_MICCAI17,Roy_ISBI17,Roy_DICTA16,Tennakoon_OMIA16} have shown moderate success for medical image classification. However, they have limited relevance for segmentation since they do not model geometric relation between different organs and most augmentation approaches do not differentiate between normal and diseased samples.  
Experiments in Section~\ref{expt:dis} show segmentation methods trained on normal subject images (Figure~\ref{fig:OCT1} (a) ) are not  equally effective for diseased cases due to significant shape changes between the two types. Hence there is a need for augmentation methods that consider the geometric relation between different anatomical regions and generate distinct images for diseased and normal cases.
Another limitation of current augmentation approaches is that they do not incorporate diversity in a principled manner. In \cite{Mahapatra_MICCAI2018,Sedai_OMIA16,Mahapatra_MLMI16,Sedai_EMBC16,Mahapatra_EMBC16,Mahapatra_MLMI15_Optic,Mahapatra_MLMI15_Prostate,Mahapatra_OMIA15} shape mask was incorporated manually for image generation, which is not practical and may lead to unrealistic deformations.

\section{Related Work}
\label{sec:prior}

\subsection{Deep Models for Retinal OCT Segmentation}

One of the first works to use multi-scale convolutional neural nets (CNNs) on OCT images \cite{RT36} employed patch-based voxel classification for detecting intraretinal fluid (IRF) and subretinal fluid (SRF) in fully supervised and weakly supervised settings. Fully convolutional neural nets and U-nets were used in \cite{RT42,RT44,MahapatraISBI15_Optic,MahapatraISBI15_JSGR,MahapatraISBI15_CD,KuangAMM14,Mahapatra_ABD2014,Schuffler_ABD2014,MahapatraISBI_CD2014} to segment IRF, and in \cite{RT45,MahapatraMICCAI_CD2013,Schuffler_ABD2013,MahapatraProISBI13,MahapatraRVISBI13,MahapatraWssISBI13,MahapatraCDFssISBI13,MahapatraCDSPIE13} to segment both the retinal layers and the fluid. Explicit fluid segmentation methods such as \cite{RT23,MahapatraABD12,MahapatraMLMI12,MahapatraSTACOM12,VosEMBC,MahapatraGRSPIE12,MahapatraMiccaiIAHBD11,MahapatraMiccai11} also achieve high classification performance.

\subsection{Data Augmentation (DA)}
While conventional augmentation approaches are easy to implement and generate a large database, their capabilities are limited in inducing data diversity. They are also sensitive to parameter values \cite{Zhao19_25,MahapatraMiccai10,MahapatraICIP10,MahapatraICDIP10a,MahapatraICDIP10b,MahapatraMiccai08,MahapatraISBI08,MahapatraICME08}, variation in image resolution, appearance and quality \cite{Zhao19_45,MahapatraICBME08_Retrieve,MahapatraICBME08_Sal,MahapatraSPIE08,MahapatraICIT06}.

Recent DL based methods trained with synthetic images outperform those trained with standard DA over classification and segmentation tasks. Antoniou et al. \cite{DAGAN,sZoom_Ar,CVIU_Ar,AMD_OCT,GANReg1_Ar,PGAN_Ar,Haze_Ar} proposed DAGAN for image generation in  few shot learning systems. Bozorgtabar et al. \cite{bozorgtabar2019syndemo} used GAN objective for domain transformation by aligning feature distribution of target data  and source domain. Mahapatra et al.  \cite{Mahapatra_MICCAI2018,Xr_Ar,RegGan_Ar,ISR_Ar,LME_Ar,Misc,Health_p} used conditional GAN (cGAN) for generating informative synthetic chest Xray images conditioned on a perturbed input mask.
GANs have also been used for generating synthetic retinal images \cite{ZhaoMIA2018,Pat2,Pat3,Pat4,Pat5,Pat6,Pat7} and brain magnetic resonance images (MRI) \cite{han2018gan,ShinSASHIMI2018}, facial expression analysis \cite{Behzad_PR2020}, for super resolution \cite{SRGAN,MahapatraMICCAI_ISR,rad2019srobb}, image registration \cite{Mahapatra_PR2020,Mahapatra_ISBI19,Mahapatra_MLMI18} and generating higher strength MRI from their low strength acquisition counterparts \cite{GAN_MI_Rev}. % 
 Generated images have implicit variations in intensity distribution but there is no explicit attempt to model attributes such as shape variations that are important to capture different conditions across a population. 
 Milletari et al. \cite{Zhao19_51,Pat8,Pat9,Pat10,Pat11,Pat12,Pat13} augmented medical images with simulated anatomical variations but demonstrate varying performance based on transformation functions and parameter settings.
%------------------------------------

\subsection{Image Generation Using Uncertainty}
 
 Kendall et al. \cite{PhS2} used approximate Bayesian inference for parameter uncertainty estimation in scene understanding, but did not capture complex correlations between different labels. Lakshminarayanan et al. \cite{PhS5} proposed a method to generate different samples using an ensemble of $M$ networks while Rupprecht et al. \cite{PhS7} presented a single network with $M$ heads for image generation. Sohn et al. \cite{PhS8} proposed a method based on conditional variational autoencoders (cVAE) to model segmentation masks, which improves the quality of generated images. In probabilistic UNet \cite{PhS4}, cVAE is combined with UNet \cite{Unet} to generate multiple segmentation masks, although with limited diversity since randomness is introduced at highest resolution only. Baumgartner et al. \cite{PhiSeg} introduced a framework to generate images with a greater diversity by injecting randomness at multiple levels.

\subsection{Our Contribution}

Based on the premise that improved data augmentation yields better segmentation performance in a DL system, we hypothesize that improved generation of synthetic images is possible  by considering the intrinsic relationships between shape and geometry of anatomical structures \cite{ShapeSim}.  In this paper we present a Geometry-Aware Shape Generative Adversarial Network (GeoGAN) that learns to generate plausible images of the desired anatomy (e.g., retinal OCT images) while preserving learned relationships between geometry and shape. We make the following contributions: 
 \begin{enumerate}
     \item Incorporating geometry information contributes to generation of realistic and qualitatively different medical images and \textbf{shapes} compared to standard DA. Other works such as \cite{Mahapatra_MICCAI2018,ZhaoMIA2018} do not incorporate this geometric relationship between anatomical parts.
    
     \item Use of uncertainty sampling and conditional shape generation on class labels to introduce diversity in the mask generation process. Compared to previous methods  we introduce diversity at different stages (different from \cite{Mahapatra_MICCAI2018,ZhaoMIA2018,PhS4}) and introduce an auxiliary classifier (different from \cite{PhiSeg,PhS8} ) for improving the quality and accuracy of generated images.
 \end{enumerate}

\section{Method}
\label{sec:method}

\begin{figure*}[h]
 \centering
\begin{tabular}{c}
\includegraphics[height=8cm, width=15cm]{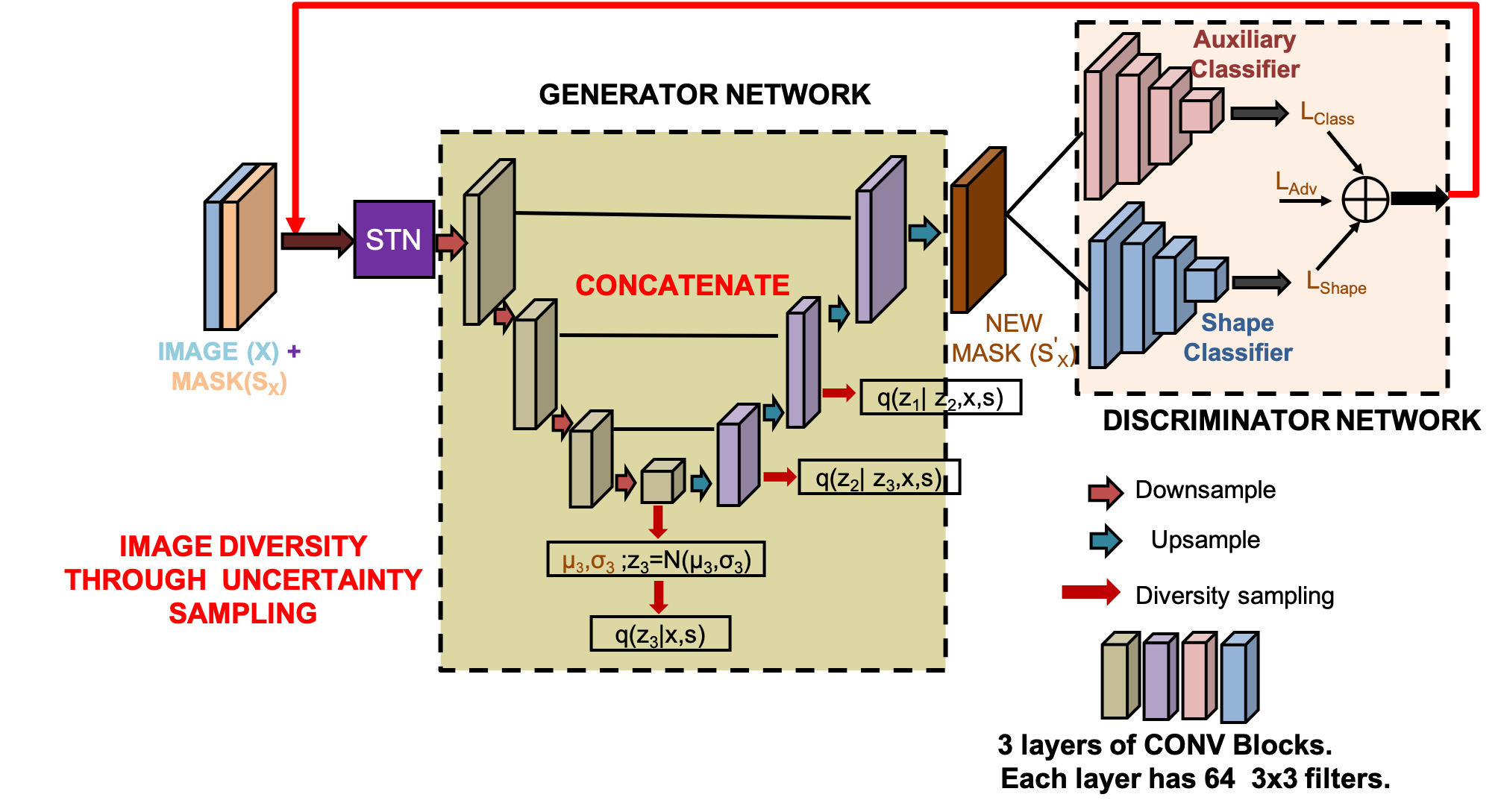}  \\
\end{tabular}
\caption{Overview of the steps in the training stage of our method. The images ($X$) and corresponding segmentation masks ($S_X$) are input to a STN whose output is fed to the generator network. Generator network is based on UNet architecture, and diversity through uncertainty sampling is injected at different levels. The generated mask $S_X^{'}$ is fed to the discriminator which evaluates its accuracy based on $L_{class}$, $L_{shape}$ and $L_{adv}$. The provided feedback is used for weight updates to obtain the final model.}
\label{fig:workflow}
\end{figure*}

\begin{figure}[h]
 \centering
\begin{tabular}{c}
\includegraphics[height=3.1cm, width=7.9cm]{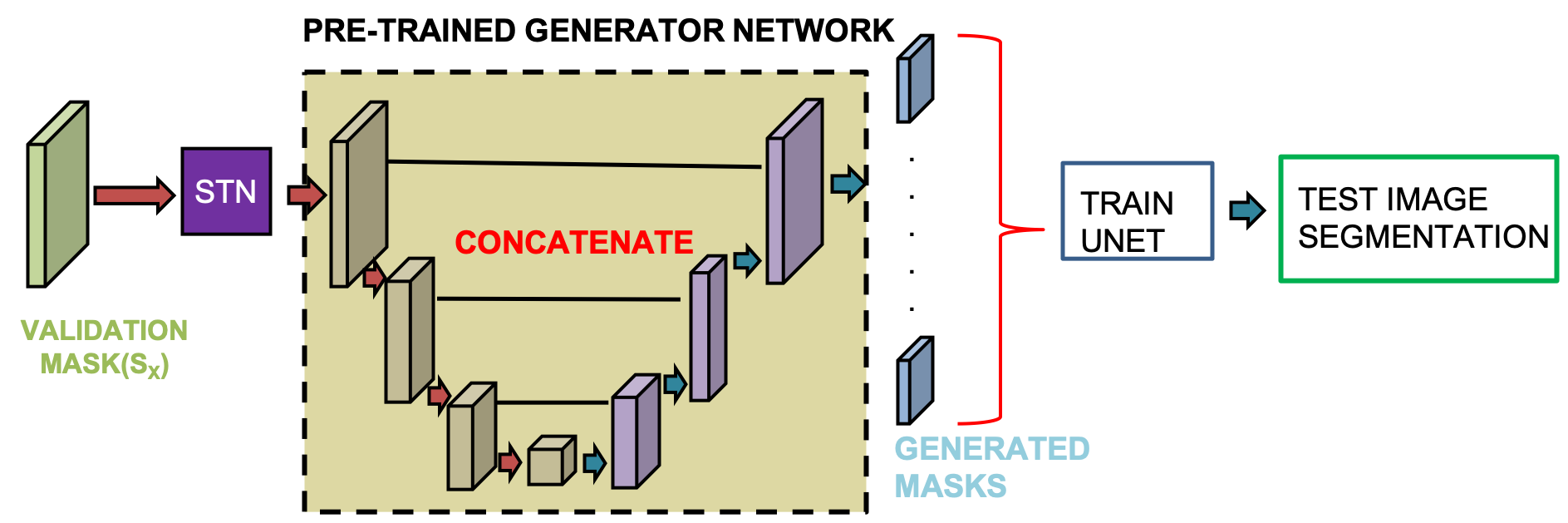}  \\
\end{tabular}
\caption{Depiction of mask generation. The trained generator network is used on validation set base images to generate new images that are used to train a segmentation network (UNet or Dense UNet). The model then segments retinal layers from test images.}
\label{fig:workflow2}
\end{figure}

Our augmentation method: 1) models geometric relationship between multiple segmentation labels; 2) preserves disease class label of original image to learn disease specific appearance and shape characteristics; and 3) introduces diversity in the image generation process through uncertainty sampling. Figure~\ref{fig:workflow} shows the training workflow using a modified UNet based generator network. The set of images and segmentation masks are used to train the generator while the discriminator provides feedback to improve the generator output. Figure~\ref{fig:workflow2} depicts generation of synthetic images from the validation image set and their subsequent use in training a UNet for image segmentation at test time.

\subsection{Geometry Aware Shape Generation}

Let us denote an input image as $x$, the corresponding manual segmentation masks as  $s_x$ and the disease class label of $x$ as $l_x$. Our method learns to generate a new image and segmentation label map from a base image and its corresponding manual mask.  
  The first stage is a spatial transformer network (STN) \cite{STN} that transforms the base mask to a new shape with different attributes of location, scale and orientation. 
 The transformations used to obtain new segmentation mask $s_x^{'}$ are applied to $x$ to get corresponding transformed image $x^{'}$.
Since the primary aim of our approach is to learn contours and other shape specific information of anatomical regions, a modified UNet architecture as the generator network effectively captures hierarchical information of shapes. It also makes it easier to introduce diversity at different levels of image abstraction.

The generator $\textbf{G}_g$ takes input $\textbf{s}_x$ and a desired label vector of output mask $c_g$ to output an affine transformation matrix $\textbf{A}$ via a STN, i.e., $\textbf{G}_g$($\textbf{s}_x, c_g) = \textbf{A}$.  
$\textbf{A}$ is used to generate  $s_x^{'}$  and  $x^{'}$. The discriminator $\textbf{D}_{class}$ determines whether output image preserves the desired label $c_g$ or not. 
The discriminator $\textbf{D}_g$ is tasked with ensuring that the generated masks and images are realistic. Let the minimax criteria between $\textbf{G}_g$ and $\textbf{D}_g$ be $\min_{\textbf{G}_g} \max_{\textbf{D}_g} \textbf{L}_g(\textbf{G}_g,\textbf{D}_g)$. The loss function $\textbf{L}_g$ has three components
 \begin{equation}
     L_g=L_{adv} + \lambda_1{L}_{class} + \lambda_2{L}_{shape}
 \label{eqn:Tloss}
 \end{equation}
 where 1) $\textbf{L}_{adv}$ is an adversarial loss to ensure $\textbf{G}_{g}$ outputs realistic deformations; 2) $\textbf{L}_{class}$ ensures generated image has characteristics of the target output class label (disease or normal); and 3) $\textbf{L}_{shape}$ ensures new masks have realistic shapes. $\lambda_1,\lambda_2$ balance each term's contribution. 

\paragraph{\textbf{Adversarial loss}}- $\textbf{L}_{adv}(\textbf{G}_g,\textbf{D}_g)$: The STN outputs $\widetilde{A}$, a prediction for $\textbf{A}$ conditioned on $\textbf{s}_x$ and a new semantic map $\textbf{s}_x \oplus \widetilde{A}(\textbf{s}_x)$ is generated. % using  composing a transformed box onto the input.
 $\textbf{L}_{adv}$ is  defined as:
% %
\begin{equation}
\begin{split}
L_{adv}(G_g,D_g) =\mathbb{E}_x\left[\log D_{g} (\textbf{s}_x \oplus \widetilde{A}(\textbf{s}_x)) \right] \\
+ \mathbb{E}_{\textbf{s}_x} \left[\log (1-D_{g} (\textbf{s}_x \oplus \widetilde{A}(\textbf{s}_x))) \right] ,
\end{split}
\label{eq:D2}
\end{equation}
% %

\paragraph{Classification Loss}- $\textbf{L}_{class}$: The affine transformation $\textbf{A}$ is applied to the base image $\textbf{x}$ to obtain  the generated image $\textbf{x}^{'}$. We add an auxiliary classifier when optimizing both $\textbf{G}_g$ and $\textbf{D}_g$ and define the classification loss as,
\begin{equation}
L_{class} = \mathbb{E}_{\textbf{x}^{'},c_g} [-\log D_{class}(c_g|x')],
\label{eq:sgan2}
\end{equation}
where the term $D_{class}(c_g|x')$ represents a probability distribution over classification labels computed by $D$. 

\paragraph{Shape Loss}-$\textbf{L}_{shape}$:
We intend to preserve the relative geometric arrangement between the different labels. 
The generated mask has regions with different assigned segmentation labels because the base mask (from which the image was generated) already has labeled layers. Let us denote by $s_i$ the image region (or pixels) in $\textbf{s}_x$ assigned label $i$. Consider another set of pixels, $s_j$, assigned label $j$. We calculate $P_{shape}(l_i|s_j,s_i)$, which is,  given regions $s_i,s_j$, the pairwise probability of $s_i$ being label $i$. 
If $n$ denotes the  total number of labels, for every label $i$ we calculate the $n-1$ such probability values and repeat it for all $n$ labels. Thus 
\begin{equation}
L_{shape} = \frac{1}{n\times(n-1)} \sum_{i,j}^{i\neq j} P_{shape} ; ~ (i,j) \in \{1,\cdots,n\}
\label{eq:sgan3}
\end{equation}
The probability value is determined from a pre-trained modified VGG16 architecture to compute $L_{shape}$ where the input has two separate maps corresponding to the label pair. Each map's foreground has only the region of the corresponding label and other labels considered background.  The conditional probability between the pair of label maps enables the classifier to implicitly capture geometrical relationships and volume information without the need to define explicit features. The geometric relation between different layers will vary for disease and normal cases, which is effectively captured by our approach.

\subsection{Sample Diversity From Uncertainty Sampling}

The generated mask $s_x'$ is obtained by fusing $L$ levels of the generator $G_g$ (as shown in Figure~\ref{fig:workflow}), each of which is associated with a latent variable $z_l$. We use probabilistic uncertainty sampling to model conditional distribution of segmentation masks and use separate latent variables at multi-resolutions to factor inherent uncertainties.
The hierarchical approach introduces diversity at different stages and influences different features (e.g., low level features at the early layers and abstract features in the later layers). Denoting the generated mask as $s$ for simplicity, we obtain conditional distribution $p(s|x)$ for $L$ latent levels as:
\begin{equation}
\begin{split}
p(s|x) = \int p(s|z_1 , \cdots, z_L )p(z_1 |z_2 , x) \cdots \\ 
     p(z_{L-1} |z_L , x)p(z_L |x) dz_1, \cdots, dz_L  
\end{split}
\label{eq:prob1}
\end{equation}

Latent variable $z_l$  models diversity at resolution  $2^{-l+1}$ of the original image (e.g. $z_1$ and $z_3$ denote the original and $1/4$ image resolution). 
A variational approximation $q(z|s, x)$ approximates the posterior distribution $p(z|s,x)$  where $z=\{z_1 , . . . , z_L\}$. $ \log p(s|x) = L(s|x) + KL(q(z|s, x)||p(z|s, x))$, where $L$ is the evidence lower bound, and $KL(.,.)$ is the Kullback-Leibler divergence. %Since $KL(·, ·) \geq 0$, $L$ is a lower bound on the conditional log probability when the approximation $q$ exactly matches the posterior. 
 The prior and posterior distributions are parameterized as normal distributions $\mathcal{N}(z|\mu,\sigma)$. %Thus, we define

Figure~\ref{fig:workflow}  shows example implementation for $L=3$. We use $6$ resolution levels and $L = 4$ latent levels.
Figure~\ref{fig:workflow}  shows the latent variables $z_l$ forming skip connections in a UNet architecture such that information between the image and segmentation output goes through a sampling step. The latent variables \emph{are not mapped} to a 1-D vector to preserve the structural  relationship between them, and this substantially improves segmentation accuracy. 
$z_l$'s dimensionality is $r_x 2^{-l+1} \times r_y 2^{ -l+1}$, where $r_x$ , $r_y$ are image dimensions. %$z_l$  models image data at $2^{-l+1}$ of original resolution due to downsampling operations and transmits the learned representation to the latent space embedding above ($z_{l-1}$). 

%
%

% this file describes the experiments and results for the paper

\section{Experimental Results}
\label{sec:expt}

\subsection{Dataset Description}
We apply our method to OCT images since retinal disease leads to significant change of retinal layers, while changes due to disease in other modalities, such as Xray or MRI, are not so obvious for mildly severe cases. Moreover, in retinal OCT there is greater interaction between different layers (segmentation labels) which is a good use case to demonstrate the effectiveness of our attempt to model the geometric relation between different anatomical regions. The publicly available RETOUCH challenge dataset \cite{Retouch} is used for our experiments. It has images of the following pathologies: 1) \textbf{Intraretinal Fluid (IRF)}: contiguous fluid-filled spaces containing columns of tissue; 2) \textbf{Subretinal Fluid (SRF)}: accumulation of a clear or lipid-rich exudate in the subretinal space; 3) \textbf{Pigment Epithelial Detachment (PED)}: detachment of the retinal pigment epithelium (RPE) along with the overlying retina from the remaining Bruch’s membrane (BM) due to the accumulation of fluid or material in sub-RPE space. It is common for age related macular degeneration (AMD). 

OCT volumes were acquired with spectral-domain SD-OCT devices from three different vendors: Cirrus HD-OCT (Zeiss Meditec), Spectralis (Heidelberg Engineering), and T-1000/T-2000 (Topcon). There were $38$ pathological OCT volumes from each vendor. 
Each Cirrus OCT consists of $128$ B-scans of $512 \times 1024$ pixels. Each Spectralis OCT had $49$ B-scans with $512 \times 496$ pixels and each Topcon OCT has $128$ B-scans of $512\times885$ (T-2000) or $512\times650$ (T-1000) pixels. All OCT volumes cover a macular area of $6\times6$ mm$^{2}$ with axial resolutions of: $2\mu$m (Cirrus), $3.9\mu$m (Spectralis), and $2.6/3.5 \mu$m (Topcon T-2000/T-1000). 
We use an additional dataset of $35$ normal subjects derived equally ($12,12,11$) from the three device types who had no incidence of retinal disease.
The training set consists of $90$ OCT volumes, with
$24, 24,$ and $22$ diseased volumes acquired with Cirrus, Spectralis, and
Topcon, respectively, with an extra $20$ normal subjects ($7,7,6$ from each device). The test set has $57$ volumes, 14 diseased volumes from each device vendor and $15$ normal subjects ($5$ from each device type). The distribution of different fluid pathologies (IRF, SRF, PED) and diseases (AMD, RVO) is almost equal in the training and test set.

The total number of images are as follows: $9071$ training images (2D scans of the volume) - $7064$ diseased and $2007$ normal; $5795$ test images- $4270$ diseased and $1525$ normal. Segmentation layers and fluid regions (in pathological images) were manually annotated in each of the ($9071+5795= 14866)$ B-scans.
Manual annotations were performed by $4$ graders and the final annotation was based on consensus.

\subsection{Experimental Setup, Baselines and Metrics}

Our method  has the following steps: 1) Split the dataset into training ($60\%$), validation ($20\%$), and test ($20\%$) folds such that images of any patient are in one fold only. 2) Use training images to train the image generator. 3) Generate shapes from the validation set and train UNet segmentation network \cite{Unet} on the generated images. 4) Use trained UNet to segment test images. 5) Repeat the above steps for different data augmentation methods. 
%a NVIDIA Titan X GPU having $12$ GB RAM, 
 We trained all models using Adam optimiser \cite{Adam} with a learning rate of $10^{-3}$ and batch-size of $12$. Batch-normalisation was used. % .
 The values of parameters $\lambda_1$ and $\lambda_2$ in Eqn.~\ref{eqn:Tloss} were set by a detailed grid search on a separate dataset of $18$ volumes ($6$ from each device) that was not used for training or testing. They were varied between $[0,1]$ in steps of $0.05$ by fixing $\lambda_1$ and varying $\lambda_2$ for the whole range. This was repeated for all values of $\lambda_1$. The best segmentation accuracy was obtained for $\lambda_1=0.9$ and $\lambda_2=0.95$, which were our final parameter values.

We denote our method as  GeoGAN (Geometry Aware GANs), and compare it's performance against other methods such as: 1) rotation, translation and scaling (denoted as DA-Data Augmentation); 2) $DAGAN$ - data augmentation GANs of \cite{DAGAN}; 3) $cGAN$ - the conditional GAN based method of \cite{Mahapatra_MICCAI2018}; and 4) $Zhao$- the atlas  registration method of \cite{Zhao_CVPR2019}.
%
%  %
Segmentation performance is evaluated in terms of Dice Metric (DM) \cite{Zhao19_23} and Hausdorff Distance (HD) \cite{HD}. DSC of $1$ indicates perfect overlap and $0$ indicates no overlap, while lower values of HD (in mm) indicate better segmentation performance.

\paragraph{Algorithm Baselines.}

The following variants of our method were used for ablation studies: 
\begin{enumerate}
    \item  GeoGAN$_{noL_{class}}$- GeoGAN without classification loss (Eqn.\ref{eq:sgan2}).
    \item GeoGAN$_{noL_{shape}}$- GeoGAN without shape relationship  modeling term (Eqn.\ref{eq:sgan3}).
    \item GeoGAN$_{NoSamp}$ - GeoGAN without  uncertainty sampling for injecting diversity to determine sampling's relevance to the final network performance. 
    %The original and upsampled images are directly connected without any sampling step.
    %
    %
    \item  GeoGAN$_{L_{class}}$ - GeoGAN using classification loss (Eqn.\ref{eq:sgan2}) and adversarial loss (Eqn.\ref{eq:D2}) to determine $L_{class}$'s  relevance to GeoGAN's performance.
    \item GeoGAN$_{L_{shape}}$ - GeoGAN using shape loss (Eqn.\ref{eq:sgan3}) and adversarial loss (Eqn.\ref{eq:D2}) to determine $L_{shape}$'s contribution to GeoGAN's performance.
    \item GeoGAN$_{Samp}$ - GeoGAN using only uncertainty sampling and adversarial loss (Eqn.\ref{eq:D2}). This baseline quantifies the contribution of sampling to the image generation process. 
\end{enumerate}

\subsection{Segmentation Results And Analysis}
\label{expt:seg}

\begin{figure*}[h]
\centering
\begin{tabular}{ccccccc}
\includegraphics[height=1.8cm, width=2.1cm]{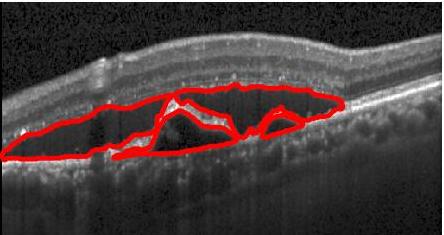} & 
\includegraphics[height=1.8cm, width=2.1cm]{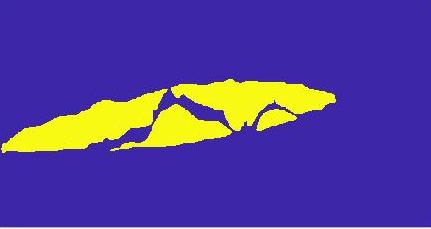} & 
\includegraphics[height=1.8cm, width=2.1cm]{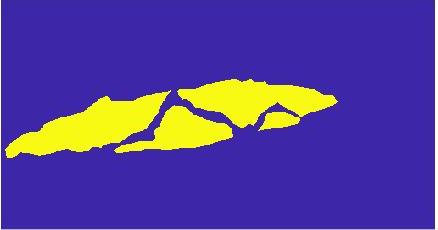} & 
\includegraphics[height=1.8cm, width=2.1cm]{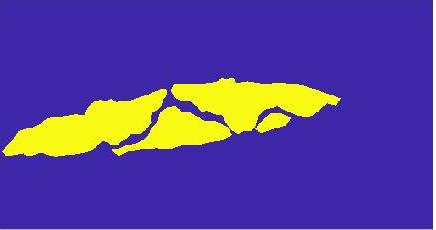} & 
\includegraphics[height=1.8cm, width=2.1cm]{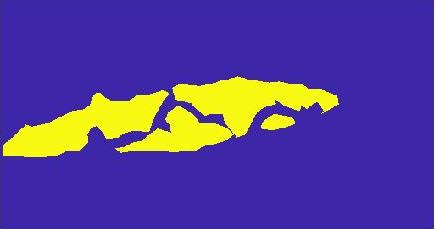} & 
\includegraphics[height=1.8cm, width=2.1cm]{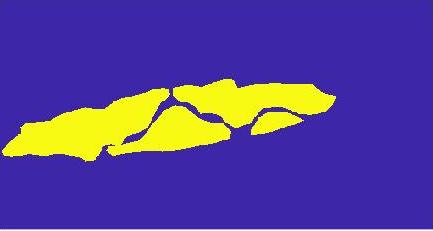} & 
\includegraphics[height=1.8cm, width=2.1cm]{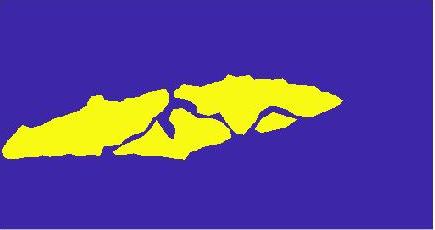} \\
(a) & (b) & (c) & (d) & (e) & (f) & (g) \\
\end{tabular}
\caption{Segmentation results on the RETOUCH challenge dataset for (a) cropped image with manual segmentation mask (red contour); Segmentation masks by (b) ground truth (manual); (c) GeoGAN; (d) Zhao \cite{Zhao_CVPR2019}; (e) $DAGAN$; (f) $cGAN$ and (g) conventional $DA$.}
\label{fig:segout1}
\end{figure*}

We hypothesize that a good image augmentation method should capture the different complex relationships between the anatomies and the generated images leading to the improvement in segmentation accuracy. 
Average DSC for \textbf{pathological images} from all device types are reported in Table~\ref{tab:OCTSeg} for the RETOUCH test dataset. % 
Figure~\ref{fig:segout1} shows the segmentation results using a UNet trained on images from different methods. Figure~\ref{fig:segout1} (a) shows the test image along with the manual mask overlayed and shown as the red contour and Figure~\ref{fig:segout1} (b) shows the manual mask. Figures~\ref{fig:segout1} (c)-(g) show, respectively, the segmentation masks obtained by GeoGAN, $Zhao~\cite{Zhao_CVPR2019}$, $DAGAN$, $cGAN$ and $DA$.

Our method outperforms baseline conventional data augmentation and other competing methods by a significant margin. Results of other methods are taken from \cite{Retouch}. GeoGAN's DSC of $0.906$ is higher than the DSC value ($0.87$) of the best performing method (obtained on the Spectralis images of the datasaet).  While GeoGAN's average performance is equally good across all three device images, the competing methods rank differently for different devices.
GeoGAN's superior segmentation accuracy is attributed to it's capacity to learn geometrical relationship between different layers (through $L_{shape}$) much better than competing methods. Thus our attempt to model the intrinsic geometrical relationships between different labels could generate superior quality masks. %

In a separate experiment we train GeoGAN with images of one device and segment images of the other devices, and repeat for all device types. The average DSC value was $0.893$, and HD was $8.6$ mm. The decrease in performance compared to GeoGAN in Table~\ref{tab:OCTSeg} is expected since the training and test images are from different devices. However we still do better than $Zhao$ \cite{Zhao_CVPR2019} and competing methods on the same dataset.

We repeat the set of experiments in Table~\ref{tab:OCTSeg} using a Dense UNet \cite{DenseUNet} instead of UNet as the segmentation network. We obtain the following average DSC values: GeoGAN -$0.917$, $Zhao~-~0.896$, $cGAN~-~0.864$, $DAGAN~-~0.834$ and $DA~-~0.802$. GeoGAN gives the best results, thus indicating it's better performance irrespective of the backbone segmentation framework.

\begin{table}[t]
 \begin{center}
\begin{tabular}{|c|c|c|c|c|c|}
\hline 
& \multicolumn{4}{|c|}{Comparison approaches} & Proposed \\ \hline
{} & {DA}  & {DAGAN}  & {cGAN} & {Zhao} & {GeoGAN} \\ 
{}& {}  & {\cite{DAGAN}} & {\cite{Mahapatra_MICCAI2018}} & {\cite{Zhao_CVPR2019}} & {}  \\ \hline
{DM} & {0.793} & {0.825} & {0.851} & {0.884} & {\textbf{0.906}} \\ 
{} & {(0.14)} & {(0.10)} & {(0.07)} & {(0.09)} & {(\textbf{0.04})} \\ \hline
{HD} & {14.3} & {12.9}  & {10.6} & {8.8} & {\textbf{7.9}} \\ 
{} & {(4.2)} & {(3.8)}  & {(3.0)} & {(3.3)} & {(\textbf{2.2})} \\ \hline
\end{tabular}
\caption{Segmentation results for \textbf{pathological} OCT images from the RETOUCH database. Mean and standard deviation (in brackets) are shown. Best results per metric is shown in bold.}% HD is in mm}
\label{tab:OCTSeg}
\end{center}
\end{table}

\begin{figure}[h]
\centering
\begin{tabular}{ccc}
\includegraphics[height=1.8cm, width=2.4cm]{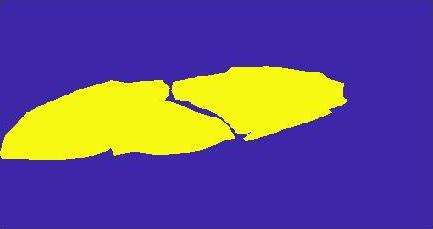} & 
\includegraphics[height=1.8cm, width=2.4cm]{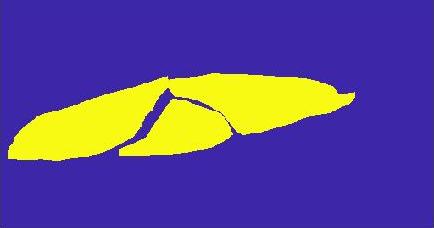} & 
\includegraphics[height=1.8cm, width=2.4cm]{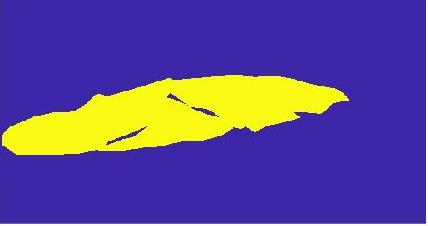} \\
(a) & (b) & (c) \\
\includegraphics[height=1.8cm, width=2.4cm]{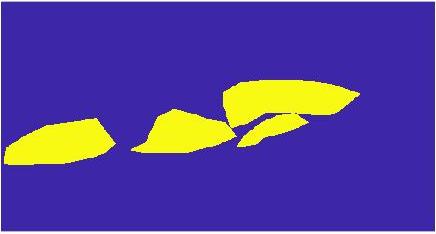} & 
\includegraphics[height=1.8cm, width=2.4cm]{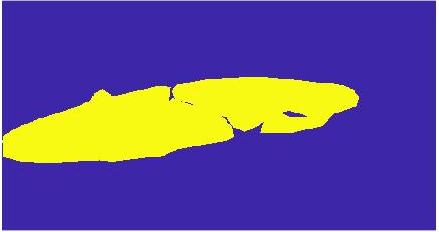} & 
\includegraphics[height=1.8cm, width=2.4cm]{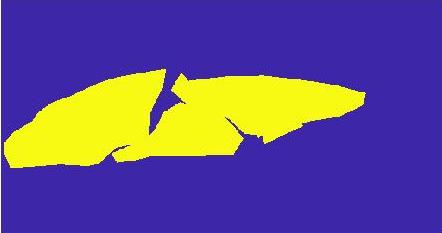}  \\
 (d) & (e) & (f) \\
\end{tabular}
\caption{Ablation study results for: (a) $GeoGAN_{noL_{shape}}$; (b) $GeoGAN_{noL_{cls}}$; (c) $GeoGAN_{noSamp}$; (d) $GeoGAN_{onlyL_{shape}}$; (e) $GeoGAN_{onlyL_{cls}}$; (f) $GeoGAN_{onlySamp}$. HD is in mm.}
\label{fig:segout1_Abl}
\end{figure}

\paragraph{Ablation Studies.}
Table~\ref{tab:OCTSeg2} shows the segmentation results for different ablation studies. Figure~\ref{fig:segout1_Abl} shows the segmentation mask obtained by different baselines for the same image shown in Figure~\ref{fig:segout1} (a). The segmentation outputs are quite different from the ground truth and the one obtained by GeoGAN. In some cases the normal regions in the layers are included as pathological area, while parts of the fluid region are not segmented as part of the pathological region. Either case is undesirable for disease diagnosis and quantification. Thus, different components of our cost functions are integral to the method's performance and excluding one or more of classification loss, geometric loss and sampling loss adversely affects segmentation performance.

\begin{table}[h]
 \begin{center}
\begin{tabular}{|c|c|c|c|}
\hline 
% & \multicolumn{4}{|c|}{Comparison approaches}  \\ \hline
{} & {GeoGAN}  & {GeoGAN}  & {GeoGAN} \\ 
{} & {$_{noL_{cls}}$}  & {$_{noL_{shape}}$}  & {$_{noSamp}$}\\ \hline
{DM} & {0.867(0.07)} & {0.864(0.09)} & {0.862(0.09)} \\ \hline
{HD} & {9.4(3.0)} & {9.5(3.3)}  & {9.9(3.2)}   \\ \hline
\hline
{} & {GeoGAN}  & {GeoGAN}  & {GeoGAN} \\ 
{} & {$_{onlyL_{cls}}$}  & {$_{onlyL_{shape}}$}  & {$_{onlySamp}$}\\ \hline
{DM} & {0.824(0.08)} & {0.825(0.07)} & {0.818(0.06)} \\ \hline
{HD} & {11.2(2.9)} & {11.1(3.0)}  & {12.5(2.8)}   \\ \hline
\end{tabular}
\caption{Mean and standard deviation (in brackets) of segmentation results from ablation studies on \textbf{pathological} OCT images from the RETOUCH database. HD is in mm.}
\label{tab:OCTSeg2}
\end{center}
\end{table}

\subsection{Realism of Synthetic Images}
\begin{figure}
\centering
\begin{tabular}{ccc}
\includegraphics[height=1.9cm, width=2.4cm]{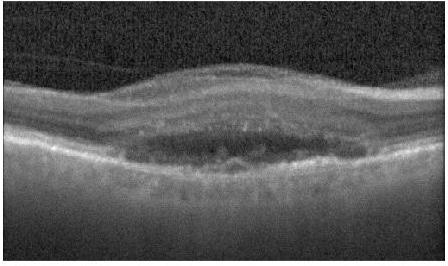} &
\includegraphics[height=1.9cm, width=2.4cm]{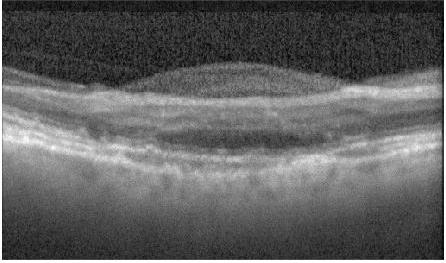} &
\includegraphics[height=1.9cm, width=2.4cm]{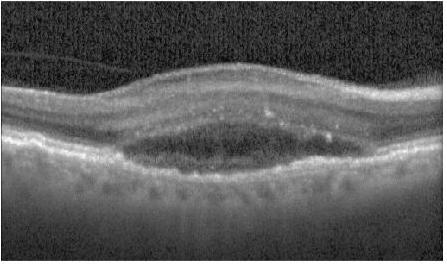} \\
(a) & (b) & (c) \\
\includegraphics[height=1.9cm, width=2.4cm]{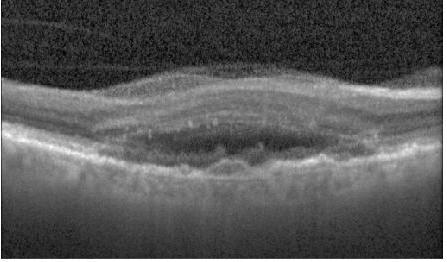} &
\includegraphics[height=1.9cm, width=2.4cm]{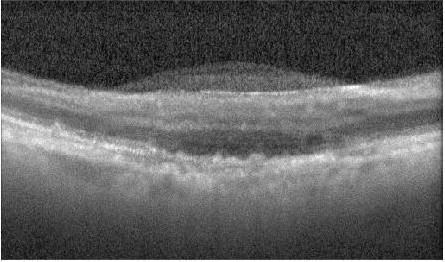} &
\includegraphics[height=1.9cm, width=2.4cm]{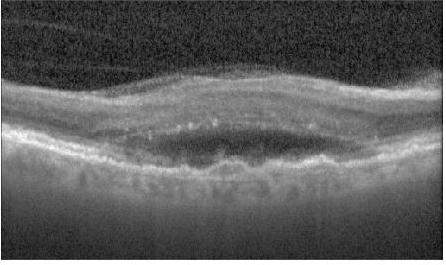}\\
 (d) & (e) & (f) \\
\end{tabular}
\caption{ Generated images for ablation study methods: (a) $GeoGAN_{noL_{cls}}$; (b) $GeoGAN_{noL_{shape}}$; (c) $GeoGAN_{noSamp}$; (d) $GeoGAN_{onlyL_{cls}}$; (e) $GeoGAN_{onlyL_{shape}}$; (f) $GeoGAN_{onlySamp}$.}
\label{fig:SynImages_Abl}
\end{figure}

Prior results show GeoGAN could generate more diverse images, which enables the corresponding UNet to show better segmentation accuracy. Figure~\ref{fig:SynImages_comp} shows examples of generated synthetic images using $GeoGAN$ and the other image generation methods except $DA$ since it involves rotation and scaling only while Figure~\ref{fig:SynImages_Abl} shows examples from the ablation models. The base image is the same in both figures. %, 
Visual examination shows GeoGAN generated images respect boundaries of adjacent layers in most cases, while other methods tend not to do so.  

Only GeoGAN and to some extent $Zhao$ generate images with consistent layer boundaries. Images generated by  other methods suffer from the following limitations: 1) tend to be  noisy; 2) multiple artifacts exposing unrealistic appearance; 3) smoothed images which distort the layer boundaries; 4) different retinal layers tend to overlap with the fluid area.  Segmentation models trained on such images will hamper their ability to produce accurate segmentations.

Two trained ophthalmologists having $4$ and $5$ years experience in examining retinal OCT images for abnormalities assessed realism of generated images. 
We present them with a common set of $500$ synthetic images from GeoGAN and ask them to classify each as realistic or not. The evaluation sessions were conducted separately with each ophthalmologist blinded to other's answers as well as the image generation model.
Results with GeoGAN show one ophthalmologist ($OPT~1$) identified $461/ 500$ ($92.2\%$) images as realistic while $OPT~2$ identified $452$ ($90.4\%$) generated images as realistic. Both of them had a high agreement with $440$ common images ($88.0\%$ -``$Both~Experts$'' in Table~\ref{tab:realism}) identified as realistic. Considering both $OPT~1$ and $OPT~2$ feedback, a total of 473 ($94.6\%$) unique images were identified as realistic (``$Atleast~1~Expert$'' in Table~\ref{tab:realism}). Subsequently, $27/500$ ($5.4\%$) of the images were not identified as realistic by any of the experts (``$No~Expert$'' in Table\ref{tab:realism}). Agreement statistics for other methods are summarized in Table~\ref{tab:realism}.

%It is important to also discuss the agreement statistics between ophthalmologists with regards to the synthetic images. 
The highest agreement between two ophthalmologists is obtained for images generated by our method. For all the other methods their difference from $GeoGAN$ is significant. Zhao et. al. \cite{Zhao_CVPR2019} has the best performance amongst them, but has agreement difference of more than $6\%$ (for ``$Atleast~1~Expert$'') compared to $GeoGAN$ ($94.6$ vs $88.2$). The numbers from Table~\ref{tab:realism} show a larger difference for the other methods, thus highlighting the importance of modeling geometric relationships in pathological region segmentation.

\begin{table}[t] 
\begin{tabular}{|c|c|c|c|}
\hline
{Agreement} & {Both} & {Atleast 1} & {No} \\ %\cline{2-4} 
{Statistics} & {Experts}  & {Expert} & {Expert}  \\ \hline
{GeoGAN} & {\textbf{88.0}~(440)}  & {\textbf{94.6}~(473)} & {\textbf{5.4}~(27)} \\ \hline
{Zhao et. al.\cite{Zhao_CVPR2019}} & {\textbf{84.8}~(424)}  & {\textbf{88.2}~(441)} & {\textbf{11.8}~(59)} \\ \hline
{cGAN (\cite{Mahapatra_MICCAI2018})} & {\textbf{83.2}~(416)}  & {\textbf{85.4}~(427)} & {\textbf{14.6}~(73)} \\ \hline
{DAGAN(\cite{DAGAN})} & {\textbf{82.2}~(411)}  & {\textbf{84.2}~(421)} & {\textbf{15.8}~(79)} \\ \hline
{DA} & {\textbf{80.4}~(402)}  & {\textbf{82.4}~(412)} & {\textbf{17.6}~(88)} \\ \hline
{GeoGAN$_{noL_{cls}}$} & {\textbf{83.6}~(418)}  & {\textbf{86.4}~(432)} & {\textbf{13.6}~(68)} \\ \hline
{GeoGAN$_{noL_{shape}}$} & {\textbf{83.0}~(415)}  & {\textbf{85.6}~(428)} & {\textbf{14.4}~(72)} \\ \hline
{GeoGAN$_{noSamp}$} & {\textbf{82.8}~(414)}  & {\textbf{85.0}~(425)} & {\textbf{15.0}~(75)} \\ \hline
{GeoGAN$_{L_{cls}}$} & {\textbf{82.2}~(411)}  & {\textbf{84.0}~(420)} & {\textbf{16.0}~(80)} \\ \hline
{GeoGAN$_{L_{shape}}$} & {\textbf{81.2}~(406)}  & {\textbf{83.4}~(417)} & {\textbf{16.6}~(83)} \\ \hline
{GeoGAN$_{Samp}$} & {\textbf{80.4}~(402)}  & {\textbf{82.8}~(414)} & {\textbf{17.2}~(86)} \\ \hline

\end{tabular}
\caption{Agreement statistics for different image generation methods amongst 2 ophthalmologists. Numbers in bold indicate agreement percentage while numbers within brackets indicate actual numbers out of $500$ patients.}
\label{tab:realism}
\end{table}

\subsection{Combining Disease And Normal Dataset}
\label{expt:dis}

Section~\ref{expt:seg} shows results of training the UNet on diseased population shapes to segment diseased shapes.  In this section we show the opposite scenario where the training was performed on normal images, the network subsequently used to generate images from the diseased base images and segment test images of a diseased population. Table~\ref{tab:MixSeg} shows the corresponding results and also for the scenario when the training images were a mix of diseased and normal population, while the test images were from the diseased population. All reported results are for the same set of test images.

\begin{figure}[h]
\centering
\begin{tabular}{ccc}
\includegraphics[height=1.8cm, width=2.4cm]{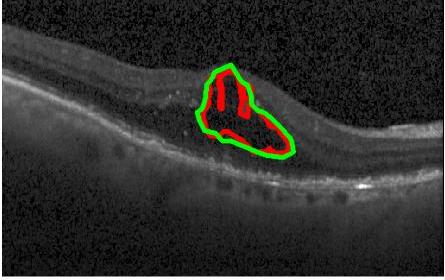} & 
\includegraphics[height=1.8cm, width=2.4cm]{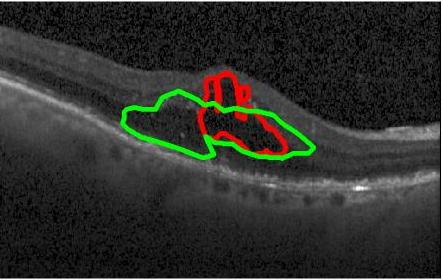} & 
\includegraphics[height=1.8cm, width=2.4cm]{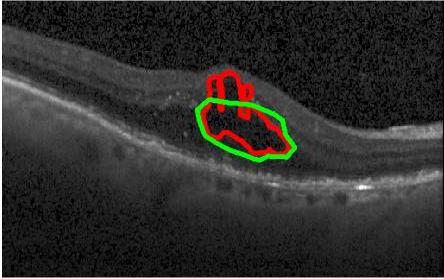}  \\
(a) & (b) & (c) \\
\end{tabular}
\caption{Segmentation results of test images for different training data sources: (a) diseased population only; (b) normal population only; (c) mix of diseased and normal population.}
\label{fig:segcomp2}
\end{figure}

Comparing them with the results in Table~\ref{tab:OCTSeg}, the superior performance of training separate networks for diseased and normal population is obvious. %Figure~\ref{fig:Segcomp2} shows an example illustrating the advantage. 
Figure~\ref{fig:segcomp2} (a) shows the segmentation output when  training and test image are from the diseased population, while Figure~\ref{fig:segcomp2} (b) shows the scenario where the training images are from the normal population while the test images are the diseased case. Red contours show the outline of the manual segmentation while the green contours show the output of our method. 
When training images are from normal population it is more challenging to segment an image from the diseased population. Inaccurate segmentation of the fluid layers can have grave consequences for subsequent diagnosis and treatment plans. Figure~\ref{fig:segcomp2} (c) shows the results when the training database is a mix of diseased and normal population, which is a more accurate representation of real world scenarios. A mixture of normal and diseased population images in the training set leads to acceptable performance. However, training a network exclusively on disease cases improves segmentation accuracy of pathological regions, which is certainly more critical than segmenting normal anatomical regions. Since it is challenging to obtain large numbers of annotated images, especially for diseased cases, our proposed image augmentation method is a significant improvement over existing methods.

\begin{table}[h]
 \begin{center}
\begin{tabular}{|c|c|c|c|c|c|}
\hline 
& \multicolumn{5}{|c|}{Train on Normal, Test on Diseased}  \\ \hline
{} & {DA}  & {DAGAN}  & {cGAN} & {\cite{Zhao_CVPR2019}} & {GeoGAN} \\ 
{}& {}  & {\cite{DAGAN}} & {\cite{Mahapatra_MICCAI2018}} & {} & {}  \\ \hline
{DM} & {0.741} & {0.781} & {0.802} & {0.821} & {\textbf{0.856}} \\ \hline
{HD} & {15.3} & {14.5}  & {13.7} & {11.3} & {\textbf{9.9}} \\ \hline
\hline
& \multicolumn{5}{|c|}{Train on Mix, Test on Diseased}  \\ \hline
{} & {DA}  & {DAGAN}  & {cGAN} & {\cite{Zhao_CVPR2019}} & {GeoGAN} \\ 
{}& {}  & {\cite{DAGAN}} & {\cite{Mahapatra_MICCAI2018}} & {} & {}  \\ \hline
{DM} & {0.762} & {0.798} & {0.820} & {0.848} & {\textbf{0.873}} \\ \hline
{HD} & {14.8} & {14.0}  & {13.2} & {10.8} & {\textbf{9.2}} \\ \hline
\end{tabular}
\caption{Segmentation results for mix of diseased and normal OCT images. Best results per metric is shown in boldface. HD is in mm.}
\label{tab:MixSeg}
\end{center}
\end{table}

\section{Conclusion}\label{sec:concl}

We propose a novel approach to generate plausible retinal OCT images by incorporating relationship between segmentation labels to guide the shape generation process.  Diversity is introduced in the image generation process through uncertainty sampling. Comparative results show that the augmented dataset from $GeoGAN$ outperforms standard data augmentation and other competing methods, when applied to segmentation of pathological regions (fluid filled areas) in retinal OCT images. We show that synergy between shape, classification and sampling terms lead to improved segmentation and greater visual agreement of experienced ophthalmologists. Each of these terms is equally important in generating realistic shapes. % e relative contributions of each term and conclude that the shape prior term makes a significant contribution to the output, while a .
 Our approach can be used for other medical imaging modalities without major changes to the workflow.
 
 Despite the good performance of our method we observe  failure cases when the base images are noisy due to inherent characteristics of the image acquisition procedure, and when the fluid areas greatly overlap with other layers. Although the second scenario is not very common, it can be critical in the medical context. In future work we aim to evaluate our method's robustness on a wide range of medical imaging modalities such as MRI, Xray, etc. 
Our method is also useful to generate realistic images for educating  clinicians, where targeted synthetic images (e.g. generation of complex cases, or disease mimickers) can be used to speed-up training. Similarly, the proposed approach could be used in quality control of deep learning systems to identify potential weaknesses through targeted high-throughput synthetic image generation and testing.

% {\small
% \bibliographystyle{ieee}
% \bibliography{CVPR2020_GeoGAN}
% }

{\small
\bibliographystyle{ieee_fullname}
\bibliography{CVPR2020_GeoGAN}

\begin{thebibliography}{100}\itemsep=-1pt

\bibitem{DAGAN}
Antreas Antoniou, Amos Storkey, and Harrison Edwards.
\newblock Data augmentation generative adversarial networks.
\newblock In {\em arXiv preprint arXiv:1711.04340,}, 2017.

\bibitem{Pat7}
P.R. Bastide, I.F. Kiral-Kornek, D. Mahapatra, S. Saha, A. Vishwanath, and
  S.~Von Cavallar.
\newblock Machine learned optimizing of health activity for participants during
  meeting times.
\newblock In {\em US Patent App. 15/426,634}, 2018.

\bibitem{Pat5}
P.R. Bastide, I.F. Kiral-Kornek, D. Mahapatra, S. Saha, A. Vishwanath, and
  S.~Von Cavallar.
\newblock Visual health maintenance and improvement.
\newblock In {\em US Patent 9,993,385}, 2018.

\bibitem{Health_p}
P.R. Bastide, I.F. Kiral-Kornek, D. Mahapatra, S. Saha, A. Vishwanath, and
  S.~Von Cavallar.
\newblock Crowdsourcing health improvements routes.
\newblock In {\em US Patent App. 15/611,519}, 2019.

\bibitem{PhiSeg}
Christian~F. Baumgartner, Kerem~C. Tezcan, Krishna Chaitanya, Andreas~M.
  Hötker, Urs~J. Muehlematter, Khoschy Schawkat, Anton~S. Becker, Olivio
  Donati, and Ender Konukoglu.
\newblock Phiseg: Capturing uncertainty in medical image segmentation.
\newblock In {\em Proc. MICCAI(2)}, pages 119--127, 2019.

\bibitem{Retouch}
Hrvoje Bogunovic and et. al.
\newblock {RETOUCH}: The retinal oct fluid detection and segmentation benchmark
  and challenge.
\newblock {\em IEEE Trans. Med. Imag.}, 38(8):1858--1874, 2019.

\bibitem{ShapeSim}
Fred~L. Bookstein.
\newblock Integration, disintegration, and self-similarity: Characterizing the
  scales of shape variation in landmark data.
\newblock {\em Evolutionary Biology}, 42(4):395--426, 2015.

\bibitem{Behzad_PR2020}
Behzad Bozorgtabar, Dwarikanath Mahapatra, and Jean-Philippe. Thiran.
\newblock Exprada: Adversarial domain adaptation for facial expression
  analysis.
\newblock {\em In Press Pattern Recognition}, 100:15--28, 2020.

\bibitem{Mahapatra_CVIU2019}
B. Bozorgtabar, D. Mahapatra, H. von Teng, A. Pollinger, L. Ebner, J-P. Thiran,
  and M. Reyes.
\newblock Informative sample generation using class aware generative
  adversarial networks for classification of chest xrays.
\newblock {\em Computer Vision and Image Understanding}, 184:57--65, 2019.

\bibitem{Mahapatra_CVIU19}
Behzad Bozorgtabar, Dwarikanath Mahapatra, Hendrik von Teng, Alexander
  Pollinger, Lucas Ebner, Jean-Philipe Thiran, and Mauricio Reyes.
\newblock Informative sample generation using class aware generative
  adversarial networks for classification of chest xrays.
\newblock {\em Computer Vision and Image Understanding}, 184:57--65, 2019.

\bibitem{CVIU_Ar}
B. Bozorgtabar, D. Mahapatra, H. von Teng, A. Pollinger, L. Ebner, J-P. Thiran,
  and M. Reyes.
\newblock Informative sample generation using class aware generative
  adversarial networks for classification of chest xrays.
\newblock In {\em arXiv preprint arXiv:1904.10781}, 2019.

\bibitem{bozorgtabar2019learn}
Behzad Bozorgtabar, Mohammad~Saeed Rad, Haz{\i}m~Kemal Ekenel, and
  Jean-Philippe Thiran.
\newblock Learn to synthesize and synthesize to learn.
\newblock {\em Computer Vision and Image Understanding}, 185:1--11, 2019.

\bibitem{Bozorgtabar_ICCV19}
B. Bozorgtabar, M.~Saeed Rad, D. Mahapatra, and J-P. Thiran.
\newblock Syndemo: Synergistic deep feature alignment for joint learning of
  depth and ego-motion.
\newblock In {\em In Proc. IEEE ICCV}, 2019.

\bibitem{bozorgtabar2019syndemo}
Behzad Bozorgtabar, Mohammad~Saeed Rad, Dwarikanath Mahapatra, and
  Jean-Philippe Thiran.
\newblock Syndemo: Synergistic deep feature alignment for joint learning of
  depth and ego-motion.
\newblock In {\em Proceedings of the IEEE International Conference on Computer
  Vision}, pages 4210--4219, 2019.

\bibitem{OCTFig}
Stephanie~J. Chiu, Xiao~T. Li, Peter Nicholas, Cynthia~A. Toth, Joseph~A.
  Izatt, and Sina Farsiu.
\newblock Automatic segmentation of seven retinal layers in sdoct images
  congruent with expert manual segmentation.
\newblock {\em Opt. Express}, 18(18):19413--19428, 2010.

\bibitem{Zhao19_23}
Lee~R. Dice.
\newblock Measures of the amount of ecologic association between species.
\newblock {\em Ecology}, 26(3):297--302, 1945.

\bibitem{Zhao19_25}
Alexey Dosovitskiy, Philipp Fischer, Jost~Tobias Springenberg, Martin
  Riedmiller, and Thomas Brox.
\newblock Discriminative unsupervised feature learning with exemplar
  convolutional neural networks.
\newblock {\em IEEE Trans. Patt. Anal. Mach. Intell.}, 38(9):1734--1747, 2016.

\bibitem{ZGe_MTA2019}
Z. Ge, D. Mahapatra, X. Chang, Z. Chen, L. Chi, and H. Lu.
\newblock Improving multi-label chest x-ray disease diagnosis by exploiting
  disease and health labels dependencies.
\newblock {\em In press Multimedia Tools and Application}, 78(22):1--14, 2019.

\bibitem{Xr_Ar}
Z. Ge, D. Mahapatra, S. Sedai, R. Garnavi, and R. Chakravorty.
\newblock Chest x-rays classification: A multi-label and fine-grained problem.
\newblock In {\em arXiv preprint arXiv:1807.07247}, 2018.

\bibitem{RT44}
G.~N. Girish, Bibhash Thakur, Sohini~Roy Chowdhurya, Abhishek~R. Kothari, and
  Jeny Rajan.
\newblock Segmentation of intra-retinal cysts from optical coherence tomography
  images using a fully convolutional neural network model.
\newblock {\em IEEE J. Biomed. Health Inform.}, 23(1):296--304, 2018.

\bibitem{goodfellow2014generative}
Ian Goodfellow, Jean Pouget-Abadie, Mehdi Mirza, Bing Xu, David Warde-Farley,
  Sherjil Ozair, Aaron Courville, and Yoshua Bengio.
\newblock Generative adversarial nets.
\newblock In {\em Advances in neural information processing systems}, pages
  2672--2680, 2014.

\bibitem{han2018gan}
Changhee Han, Hideaki Hayashi, Leonardo Rundo, Ryosuke Araki, Wataru Shimoda,
  Shinichi Muramatsu, Yujiro Furukawa, Giancarlo Mauri, and Hideki Nakayama.
\newblock Gan-based synthetic brain mr image generation.
\newblock In {\em 2018 IEEE 15th International Symposium on Biomedical Imaging
  (ISBI 2018)}, pages 734--738. IEEE, 2018.

\bibitem{huang2018auggan}
Sheng-Wei Huang, Che-Tsung Lin, Shu-Ping Chen, Yen-Yi Wu, Po-Hao Hsu, and
  Shang-Hong Lai.
\newblock Auggan: Cross domain adaptation with gan-based data augmentation.
\newblock In {\em Proceedings of the European Conference on Computer Vision
  (ECCV)}, pages 718--731, 2018.

\bibitem{STN}
Max Jaderberg, Karen Simonyan, Andrew Zisserman, and Koray Kavukcuoglu.
\newblock Spatial transformer networks.
\newblock In {\em NIPS}, pages~--, 2015.

\bibitem{PhS2}
Alex Kendall, Vijay Badrinarayanan, and Roberto Cipolla.
\newblock Bayesian segnet: Model uncertainty in deep convolutional
  encoder-decoder architectures for scene understanding.
\newblock In {\em arXiv:1511.02680}, 2015.

\bibitem{Adam}
Diederik~P. Kingma and Jimmy Ba.
\newblock Adam: A method for stochastic optimization.
\newblock In {\em arXiv preprint arXiv:1412.6980,}, 2014.

\bibitem{PhS4}
Simon A.~A. Kohl, Bernardino Romera-Paredes, Clemens Meyer, Jeffrey~De Fauw,
  Joseph~R. Ledsam, Klaus~H. Maier-Hein, S.~M.~Ali Eslami, Danilo~Jimenez
  Rezende, and Olaf Ronneberger.
\newblock A probabilistic u-net for segmentation of ambiguous images.
\newblock In {\em Proc. NIPS}, pages 6965--6975, 2018.

\bibitem{Kuanar_ICIP19}
S. Kuanar, V Athitsos, D. Mahapatra, K.R. Rao, Z. Akhtar, and D. Dasgupta.
\newblock Low dose abdominal ct image reconstruction: An unsupervised learning
  based approach.
\newblock In {\em In Proc. IEEE ICIP}, pages 1351--1355, 2019.

\bibitem{Haze_Ar}
S. Kuanar, K.R. Rao, D. Mahapatra, and M. Bilas.
\newblock Night time haze and glow removal using deep dilated convolutional
  network.
\newblock In {\em arXiv preprint arXiv:1902.00855}, 2019.

\bibitem{KuangAMM14}
H. Kuang, B. Guthier, M. Saini, D. Mahapatra, and A.~El Saddik.
\newblock A real-time smart assistant for video surveillance through handheld
  devices.
\newblock In {\em In Proc: ACM Intl. Conf. Multimedia}, pages 917--920, 2014.

\bibitem{PhS5}
Balaji Lakshminarayanan, Alexander Pritzel, and Charles Blundell.
\newblock Simple and scalable predictive uncertainty estimation using deep
  ensembles.
\newblock In {\em Proc. NIPS}, pages 6402--6413, 2017.

\bibitem{SRGAN}
Christian Ledig, Lucas Theis, Ferenc Huszar, Jose Caballero, Andrew Cunningham,
  Alejandro Acosta, Andrew Aitken, Alykhan Tejani, Johannes Totz, Zehan Wang,
  and Wenzhe Shi.
\newblock Photo-realistic single image super-resolution using a generative
  adversarial network.
\newblock {\em CoRR}, abs/1609.04802, 2016.

\bibitem{Zhao19_45}
Kelvin~K. Leung, Matthew~J. Clarkson, Johnathon~W. Bartlett, Shona Clegg,
  Clifford R.~Jack Jr, Michael~W. Weiner, Nick~C. Fox, Sebastien Ourselin, and
  A.~D.~N. Initiative.
\newblock Robust atrophy rate measurement in alzheimer’s disease using
  multi-site serial {mri}: tissue-specific intensity normalization and
  parameter selection.
\newblock {\em Neuroimage}, 50(2):516--523, 2010.

\bibitem{DenseUNet}
Xiaomeng Li, Hao Chen, Xiaojuan Qi, Qi Dou, Chi-Wing Fu, and Pheng-Ann Heng.
\newblock {H-DenseUNet}: Hybrid densely connected unet for liver and tumor
  segmentation from ct volumes.
\newblock {\em IEEE Trans. Med. Imag.}, 37(12):2663--2674, 2018.

\bibitem{LiTMI_2015}
Z. Li, D. Mahapatra, J.Tielbeek, J. Stoker, L. van Vliet, and F.M. Vos.
\newblock Image registration based on autocorrelation of local structure.
\newblock {\em IEEE Trans. Med. Imaging}, 35(1):63--75, 2016.

\bibitem{MahapatraMiccaiIAHBD11}
D. Mahapatra.
\newblock Neonatal brain mri skull stripping using graph cuts and shape priors.
\newblock In {\em In Proc: MICCAI workshop on Image Analysis of Human Brain
  Development (IAHBD)}, 2011.

\bibitem{MahapatraMLMI12}
D. Mahapatra.
\newblock Cardiac lv and rv segmentation using mutual context information.
\newblock In {\em Proc. MICCAI-MLMI}, pages 201--209, 2012.

\bibitem{MahapatraGRSPIE12}
D. Mahapatra.
\newblock Groupwise registration of dynamic cardiac perfusion images using
  temporal information and segmentation information.
\newblock In {\em In Proc: SPIE Medical Imaging}, 2012.

\bibitem{MahapatraSTACOM12}
D. Mahapatra.
\newblock Landmark detection in cardiac mri using learned local image
  statistics.
\newblock In {\em Proc. MICCAI-Statistical Atlases and Computational Models of
  the Heart. Imaging and Modelling Challenges (STACOM)}, pages 115--124, 2012.

\bibitem{MahapatraJDISkull2012}
D. Mahapatra.
\newblock Skull stripping of neonatal brain mri: Using prior shape information
  with graphcuts.
\newblock {\em J. Digit. Imaging}, 25(6):802--814, 2012.

\bibitem{MahapatraJDIGCSP2013}
D. Mahapatra.
\newblock Cardiac image segmentation from cine cardiac mri using graph cuts and
  shape priors.
\newblock {\em J. Digit. Imaging}, 26(4):721--730, 2013.

\bibitem{MahapatraJDIMutCont2013}
D. Mahapatra.
\newblock Cardiac mri segmentation using mutual context information from left
  and right ventricle.
\newblock {\em J. Digit. Imaging}, 26(5):898--908, 2013.

\bibitem{MahapatraProISBI13}
D. Mahapatra.
\newblock Graph cut based automatic prostate segmentation using learned
  semantic information.
\newblock In {\em Proc. IEEE ISBI}, pages 1304--1307, 2013.

\bibitem{MahapatraJDIJSGR2013}
D. Mahapatra.
\newblock Joint segmentation and groupwise registration of cardiac perfusion
  images using temporal information.
\newblock {\em J. Digit. Imaging}, 26(2):173--182, 2013.

\bibitem{MahapatraJDI_Cardiac_FSL}
D. Mahapatra.
\newblock Automatic cardiac segmentation using semantic information from random
  forests.
\newblock {\em J. Digit. Imaging.}, 27(6):794--804, 2014.

\bibitem{Mahapatra_LME_CVIU}
D. Mahapatra.
\newblock Combining multiple expert annotations using semi-supervised learning
  and graph cuts for medical image segmentation.
\newblock {\em Computer Vision and Image Understanding}, 151(1):114--123, 2016.

\bibitem{LME_Ar}
D. Mahapatra.
\newblock Consensus based medical image segmentation using semi-supervised
  learning and graph cuts.
\newblock In {\em arXiv preprint arXiv:1612.02166}, 2017.

\bibitem{Mahapatra_LME_PR2017}
D. Mahapatra.
\newblock Semi-supervised learning and graph cuts for consensus based medical
  image segmentation.
\newblock {\em Pattern Recognition}, 63(1):700--709, 2017.

\bibitem{AMD_OCT}
D. Mahapatra.
\newblock Amd severity prediction and explainability using image registration
  and deep embedded clustering.
\newblock In {\em arXiv preprint arXiv:1907.03075}, 2019.

\bibitem{Misc}
D. Mahapatra, K. Agarwal, R. Khosrowabadi, and D.K. Prasad.
\newblock Recent advances in statistical data and signal analysis: Application
  to real world diagnostics from medical and biological signals.
\newblock In {\em Computational and mathematical methods in medicine}, 2016.

\bibitem{MahapatraGAN_ISBI18}
D. Mahapatra, B. Antony, S. Sedai, and R. Garnavi.
\newblock Deformable medical image registration using generative adversarial
  networks.
\newblock In {\em In Proc. IEEE ISBI}, pages 1449--1453, 2018.

\bibitem{ISR_Ar}
D. Mahapatra and B. Bozorgtabar.
\newblock Retinal vasculature segmentation using local saliency maps and
  generative adversarial networks for image super resolution.
\newblock In {\em arXiv preprint arXiv:1710.04783}, 2017.

\bibitem{PGAN_Ar}
D. Mahapatra and B. Bozorgtabar.
\newblock Progressive generative adversarial networks for medical image super
  resolution.
\newblock In {\em arXiv preprint arXiv:1902.02144}, 2019.

\bibitem{Mahapatra_CMIG2019}
D. Mahapatra, B. Bozorgtabar, and R. Garnavi.
\newblock Image super-resolution using progressive generative adversarial
  networks for medical image analysis.
\newblock {\em Computerized Medical Imaging and Graphics}, 71:30--39, 2019.

\bibitem{MahapatraMICCAI_ISR}
Dwarikanath Mahapatra, Behzad Bozorgtabar, and Sajini Hewavitharanage.
\newblock Image super resolution using generative adversarial networks and
  local saliency maps for retinal image analysis.
\newblock In {\em MICCAI}, pages 382--390, 2017.

\bibitem{Mahapatra_MICCAI2018}
Dwarikanath Mahapatra, Behzad Bozorgtabar, Jean-Phillipe Thiran, and Mauricio
  Reyes.
\newblock Efficient active learning for image classification and segmentation
  using a sample selection and conditional generative adversarial network.
\newblock In {\em MICCAI}, pages 580--588, 2018.

\bibitem{Mahapatra_MICCAI17}
D. Mahapatra, S. Bozorgtabar, S. Hewavitahranage, and R. Garnavi.
\newblock Image super resolution using generative adversarial networks and
  local saliencymaps for retinal image analysis,.
\newblock In {\em In Proc. MICCAI}, pages 382--390, 2017.

\bibitem{MahapatraAL_MICCAI18}
D. Mahapatra, S. Bozorgtabar, J.-P. Thiran, and M. Reyes.
\newblock Efficient active learning for image classification and segmentation
  using a sample selection and conditional generative adversarial network.
\newblock In {\em In Proc. MICCAI (2)}, pages 580--588, 2018.

\bibitem{MahapatraRVISBI13}
D. Mahapatra and J.M. Buhmann.
\newblock Automatic cardiac rv segmentation using semantic information with
  graph cuts.
\newblock In {\em Proc. IEEE ISBI}, pages 1094--1097, 2013.

\bibitem{MahapatraTIP_RF2014}
D. Mahapatra and J.M. Buhmann.
\newblock Analyzing training information from random forests for improved image
  segmentation.
\newblock {\em IEEE Trans. Imag. Proc.}, 23(4):1504--1512, 2014.

\bibitem{MahapatraTBME_Pro2014}
D. Mahapatra and J.M. Buhmann.
\newblock Prostate mri segmentation using learned semantic knowledge and graph
  cuts.
\newblock {\em IEEE Trans. Biomed. Engg.}, 61(3):756--764, 2014.

\bibitem{MahapatraISBI15_Optic}
D. Mahapatra and J.M. Buhmann.
\newblock A field of experts model for optic cup and disc segmentation from
  retinal fundus images.
\newblock In {\em In Proc. IEEE ISBI}, pages 218--221, 2015.

\bibitem{Mahapatra_OMIA15}
D. Mahapatra and J. Buhmann.
\newblock Obtaining consensus annotations for retinal image segmentation using
  random forest and graph cuts.
\newblock In {\em In Proc. OMIA}, pages 41--48, 2015.

\bibitem{Mahapatra_MLMI15_Prostate}
D. Mahapatra and J. Buhmann.
\newblock Visual saliency based active learning for prostate mri segmentation.
\newblock In {\em In Proc. MLMI}, pages 9--16, 2015.

\bibitem{Mahapatra_SSLAL_Pro_JMI}
D. Mahapatra and J. Buhmann.
\newblock Visual saliency based active learning for prostate mri segmentation.
\newblock {\em SPIE Journal of Medical Imaging}, 3(1), 2016.

\bibitem{Pat2}
D. Mahapatra, R Garnavi, P.K. Roy, and R.B. Tennakoon.
\newblock System and method to teach and evaluate image grading performance
  using prior learned expert knowledge base.
\newblock In {\em US Patent App. 15/459,457}, 2018.

\bibitem{Pat3}
D. Mahapatra, R Garnavi, P.K. Roy, and R.B. Tennakoon.
\newblock System and method to teach and evaluate image grading performance
  using prior learned expert knowledge base.
\newblock In {\em US Patent App. 15/814,590}, 2018.

\bibitem{Pat11}
D. Mahapatra, R Garnavi, S. Sedai, and P.K. Roy.
\newblock Joint segmentation and characteristics estimation in medical images.
\newblock In {\em US Patent App. 15/234,426}, 2017.

\bibitem{Pat10}
D. Mahapatra, R Garnavi, S. Sedai, and P.K. Roy.
\newblock Retinal image quality assessment, error identification and automatic
  quality correction.
\newblock In {\em US Patent 9,779,492}, 2017.

\bibitem{Pat6}
D. Mahapatra, R Garnavi, S. Sedai, and R.B. Tennakoon.
\newblock Classification of severity of pathological condition using hybrid
  image representation.
\newblock In {\em US Patent App. 15/426,634}, 2018.

\bibitem{Pat4}
D. Mahapatra, R Garnavi, S. Sedai, and R.B. Tennakoon.
\newblock Generating an enriched knowledge base from annotated images.
\newblock In {\em US Patent App. 15/429,735}, 2018.

\bibitem{Pat8}
D. Mahapatra, R Garnavi, S. Sedai, R.B. Tennakoon, and R. Chakravorty.
\newblock Early prediction of age related macular degeneration by image
  reconstruction.
\newblock In {\em US Patent App. 15/854,984}, 2018.

\bibitem{Pat9}
D. Mahapatra, R Garnavi, S. Sedai, R.B. Tennakoon, and R. Chakravorty.
\newblock Early prediction of age related macular degeneration by image
  reconstruction.
\newblock In {\em US Patent 9,943,225}, 2018.

\bibitem{GANReg1_Ar}
D. Mahapatra and Z. Ge.
\newblock Combining transfer learning and segmentation information with gans
  for training data independent image registration.
\newblock In {\em arXiv preprint arXiv:1903.10139}, 2019.

\bibitem{Mahapatra_ISBI19}
Dwarikanath Mahapatra and Zongyuan Ge.
\newblock Training data independent image registration with gans using transfer
  learning and segmentation information.
\newblock In {\em In Proc. IEEE ISBI}, pages 709--713, 2019.

\bibitem{Mahapatra_PR2020}
Dwarikanath Mahapatra and Zongyuan Ge.
\newblock Training data independent image registration using generative
  adversarial networks and domain adaptation.
\newblock {\em In press Pattern Recognition}, 100:1--14, 2020.

\bibitem{Pat13}
D. Mahapatra, Z. Ge, and S. Sedai.
\newblock Joint registration and segmentation of images using deep learning.
\newblock In {\em US Patent App. 16/001,566}, 2019.

\bibitem{Mahapatra_MLMI18}
Dwarikanath Mahapatra, Zongyuan Ge, Suman Sedai, and Rajib Chakravorty.
\newblock Joint registration and segmentation of xray images using generative
  adversarial networks.
\newblock In {\em In Proc. MICCAI-MLMI}, pages 73--80, 2018.

\bibitem{Mahapatra_JSTSP2014}
D. Mahapatra, S. Gilani, and M.K. Saini.
\newblock Coherency based spatio-temporal saliency detection for video object
  segmentation.
\newblock {\em IEEE Journal of Selected Topics in Signal Processing.},
  8(3):454--462, 2014.

\bibitem{MahapatraTMI_CD2013}
D. Mahapatra, J.Tielbeek, J.C. Makanyanga, J. Stoker, S.A. Taylor, F.M. Vos,
  and J.M. Buhmann.
\newblock Automatic detection and segmentation of crohn's disease tissues from
  abdominal mri.
\newblock {\em IEEE Trans. Med. Imaging}, 32(12):1232--1248, 2013.

\bibitem{MahapatraISBI_CD2014}
D. Mahapatra, J.Tielbeek, J.C. Makanyanga, J. Stoker, S.A. Taylor, F.M. Vos,
  and J.M. Buhmann.
\newblock Active learning based segmentation of crohn's disease using
  principles of visual saliency.
\newblock In {\em Proc. IEEE ISBI}, pages 226--229, 2014.

\bibitem{Mahapatra_ABD2014}
D. Mahapatra, J.Tielbeek, J.C. Makanyanga, J. Stoker, S.A. Taylor, F.M. Vos,
  and J.M. Buhmann.
\newblock Combining multiple expert annotations using semi-supervised learning
  and graph cuts for crohn's disease segmentation.
\newblock In {\em In Proc: MICCAI-ABD}, 2014.

\bibitem{MahapatraJDICD2013}
D. Mahapatra, J.Tielbeek, F.M. Vos, and J.M. Buhmann.
\newblock A supervised learning approach for crohn's disease detection using
  higher order image statistics and a novel shape asymmetry measure.
\newblock {\em J. Digit. Imaging}, 26(5):920--931, 2013.

\bibitem{MahapatraISBI15_JSGR}
D. Mahapatra, Z. Li, F.M. Vos, and J.M. Buhmann.
\newblock Joint segmentation and groupwise registration of cardiac dce mri
  using sparse data representations.
\newblock In {\em In Proc. IEEE ISBI}, pages 1312--1315, 2015.

\bibitem{MahapatraICIT06}
D. Mahapatra, A. Routray, and C. Mishra.
\newblock An active snake model for classification of extreme emotions.
\newblock In {\em IEEE International Conference on Industrial Technology
  (ICIT)}, pages 2195--2199, 2006.

\bibitem{Mahapatra_EMBC16}
D. Mahapatra, P.K. Roy, S. Sedai, and R. Garnavi.
\newblock A cnn based neurobiology inspired approach for retinal image quality
  assessment.
\newblock In {\em In Proc. EMBC}, pages 1304--1307, 2016.

\bibitem{Mahapatra_MLMI16}
D. Mahapatra, P.K. Roy, S. Sedai, and R. Garnavi.
\newblock Retinal image quality classification using saliency maps and cnns.
\newblock In {\em In Proc. MICCAI-MLMI}, pages 172--179, 2016.

\bibitem{MahapatraICBME08_Retrieve}
D. Mahapatra, S. Roy, and Y. Sun.
\newblock Retrieval of mr kidney images by incorporating spatial information in
  histogram of low level features.
\newblock In {\em In 13th International Conference on Biomedical Engineering},
  2008.

\bibitem{Pat12}
D. Mahapatra, S. Saha, A. Vishwanath, and P.R. Bastide.
\newblock Generating hyperspectral image database by machine learning and
  mapping of color images to hyperspectral domain.
\newblock In {\em US Patent App. 15/949,528}, 2019.

\bibitem{MahapatraICME08}
D. Mahapatra, M.K. Saini, and Y. Sun.
\newblock Illumination invariant tracking in office environments using
  neurobiology-saliency based particle filter.
\newblock In {\em IEEE ICME}, pages 953--956, 2008.

\bibitem{MahapatraMICCAI_CD2013}
D. Mahapatra, P. Sch$\ddot{u}$ffler, J. Tielbeek, F.M. Vos, and J.M. Buhmann.
\newblock Semi-supervised and active learning for automatic segmentation of
  crohn's disease.
\newblock In {\em Proc. MICCAI, Part 2}, pages 214--221, 2013.

\bibitem{RegGan_Ar}
D. Mahapatra, S. Sedai, and R. Garnavi.
\newblock Elastic registration of medical images with gans.
\newblock In {\em arXiv preprint arXiv:1805.02369}, 2018.

\bibitem{MahapatraMiccai08}
D. Mahapatra and Y. Sun.
\newblock Nonrigid registration of dynamic renal {MR} images using a saliency
  based {MRF} model.
\newblock In {\em Proc. MICCAI}, pages 771--779, 2008.

\bibitem{MahapatraISBI08}
D. Mahapatra and Y. Sun.
\newblock Registration of dynamic renal mr images using neurobiological model
  of saliency.
\newblock In {\em Proc. ISBI}, pages 1119--1122, 2008.

\bibitem{MahapatraICBME08_Sal}
D. Mahapatra and Y. Sun.
\newblock Using saliency features for graphcut segmentation of perfusion kidney
  images.
\newblock In {\em In 13th International Conference on Biomedical Engineering},
  2008.

\bibitem{MahapatraMiccai10}
D. Mahapatra and Y. Sun.
\newblock Joint registration and segmentation of dynamic cardiac perfusion
  images using mrfs.
\newblock In {\em Proc. MICCAI}, pages 493--501, 2010.

\bibitem{MahapatraICIP10}
D. Mahapatra and Y. Sun.
\newblock An mrf framework for joint registration and segmentation of natural
  and perfusion images.
\newblock In {\em Proc. IEEE ICIP}, pages 1709--1712, 2010.

\bibitem{MahapatraICDIP10a}
D. Mahapatra and Y. Sun.
\newblock Retrieval of perfusion images using cosegmentation and shape context
  information.
\newblock In {\em Proc. APSIPA Annual Summit and Conference (ASC)}, 2010.

\bibitem{MahapatraEURASIP2010}
D. Mahapatra and Y. Sun.
\newblock Rigid registration of renal perfusion images using a neurobiology
  based visual saliency model.
\newblock {\em EURASIP Journal on Image and Video Processing.}, pages 1--16,
  2010.

\bibitem{MahapatraICDIP10b}
D. Mahapatra and Y. Sun.
\newblock A saliency based mrf method for the joint registration and
  segmentation of dynamic renal mr images.
\newblock In {\em Proc. ICDIP}, 2010.

\bibitem{MahapatraTBME2011}
D. Mahapatra and Y. Sun.
\newblock Mrf based intensity invariant elastic registration of cardiac
  perfusion images using saliency information.
\newblock {\em IEEE Trans. Biomed. Engg.}, 58(4):991--1000, 2011.

\bibitem{MahapatraMiccai11}
D. Mahapatra and Y. Sun.
\newblock Orientation histograms as shape priors for left ventricle
  segmentation using graph cuts.
\newblock In {\em In Proc: MICCAI}, pages 420--427, 2011.

\bibitem{MahapatraTIP2012}
D. Mahapatra and Y. Sun.
\newblock Integrating segmentation information for improved mrf-based elastic
  image registration.
\newblock {\em IEEE Trans. Imag. Proc.}, 21(1):170--183, 2012.

\bibitem{MahapatraABD12}
D. Mahapatra, J. Tielbeek, J.M. Buhmann, and F.M. Vos.
\newblock A supervised learning based approach to detect crohn's disease in
  abdominal mr volumes.
\newblock In {\em Proc. MICCAI workshop Computational and Clinical Applications
  in Abdominal Imaging(MICCAI-ABD)}, pages 97--106, 2012.

\bibitem{MahapatraCDFssISBI13}
D. Mahapatra, J. Tielbeek, F.M. Vos, and J.M.~Buhmann .
\newblock Crohn's disease tissue segmentation from abdominal mri using semantic
  information and graph cuts.
\newblock In {\em Proc. IEEE ISBI}, pages 358--361, 2013.

\bibitem{MahapatraCDSPIE13}
D. Mahapatra, J. Tielbeek, F.M. Vos, and J.M. Buhmann.
\newblock Localizing and segmenting crohn's disease affected regions in
  abdominal mri using novel context features.
\newblock In {\em Proc. SPIE Medical Imaging}, 2013.

\bibitem{MahapatraWssISBI13}
D. Mahapatra, J. Tielbeek, F.M. Vos, and J.M. Buhmann.
\newblock Weakly supervised semantic segmentation of crohn's disease tissues
  from abdominal mri.
\newblock In {\em Proc. IEEE ISBI}, pages 832--835, 2013.

\bibitem{MahapatraISBI15_CD}
D. Mahapatra, F.M. Vos, and J.M. Buhmann.
\newblock Crohn's disease segmentation from mri using learned image priors.
\newblock In {\em In Proc. IEEE ISBI}, pages 625--628, 2015.

\bibitem{Mahapatra_SSLAL_CD_CMPB}
D. Mahapatra, F.M. Vos, and J.M. Buhmann.
\newblock Active learning based segmentation of crohns disease from abdominal
  mri.
\newblock {\em Computer Methods and Programs in Biomedicine}, 128(1):75--85,
  2016.

\bibitem{MahapatraSPIE08}
D. Mahapatra, S. Winkler, and S.C. Yen.
\newblock Motion saliency outweighs other low-level features while watching
  videos.
\newblock In {\em SPIE HVEI.}, pages 1--10, 2008.

\bibitem{Zhao19_51}
Fausto Milletari, Nassir Navab, and Seyed-Ahmad Ahmadi.
\newblock V-net: Fully convolutional neural networks for volumetric medical im-
  age segmentation.
\newblock In {\em Proc. Int. Conf. on 3D vision}, pages 565--571, 2016.

\bibitem{nielsen2019gan}
Christopher Nielsen and Michal Okoniewski.
\newblock Gan data augmentation through active learning inspired sample
  acquisition.
\newblock In {\em Proceedings of the IEEE Conference on Computer Vision and
  Pattern Recognition Workshops}, pages 109--112, 2019.

\bibitem{rad2019srobb}
Mohammad~Saeed Rad, Behzad Bozorgtabar, Urs-Viktor Marti, Max Basler,
  Hazim~Kemal Ekenel, and Jean-Philippe Thiran.
\newblock Srobb: Targeted perceptual loss for single image super-resolution.
\newblock In {\em Proceedings of the IEEE International Conference on Computer
  Vision}, pages 2710--2719, 2019.

\bibitem{HD}
Javier Ribera, David Güera, Yuhao Chen, and Edward Delp.
\newblock Weighted hausdorff distance: A loss function for object localization.
\newblock In {\em arXiv preprint arXiv:1806.07564}, 2018.

\bibitem{Unet}
Olaf Ronneberger, Phillip Fischer, and Thomas Brox.
\newblock U-net: Convolutional networks for biomedical image segmentation.
\newblock In {\em In Proc. MICCAI}, pages 234--241, 2015.

\bibitem{RT45}
Abhijit~Guha Roy, Sailesh Conjeti, Sri Phani~Krishna Karri, Debdoot Sheet, Amin
  Katouzian, Christian Wachinger, and Nassir Navab.
\newblock Relaynet: retinal layer and fluid segmentation of macular optical
  coherence tomography using fully convolutional networks.
\newblock {\em Biomed. Opt. Express}, 8(8):3627--3642, 2017.

\bibitem{Roy_DICTA16}
P. Roy, R. Chakravorty, S. Sedai, D. Mahapatra, and R. Garnavi.
\newblock Automatic eye type detection in retinal fundus image using fusion of
  transfer learning and anatomical features.
\newblock In {\em In Proc. DICTA}, pages 1--7, 2016.

\bibitem{Roy_ISBI17}
P. Roy, R. Tennakoon, K. Cao, S. Sedai, D. Mahapatra, S. Maetschke, and R.
  Garnavi.
\newblock A novel hybrid approach for severity assessment of diabetic
  retinopathy in colour fundus images,.
\newblock In {\em In Proc. IEEE ISBI}, pages 1078--1082, 2017.

\bibitem{PhS7}
Christian Rupprecht, Iro Laina, Robert DiPietro, Maximilian Baust, Federico
  Tombari, Nassir Navab, and Gregory~D. Hager.
\newblock Learning in an uncertain world: Representing ambiguity through
  multiple hypotheses.
\newblock In {\em Proc. CVPR}, pages 3591--3600, 2017.

\bibitem{sZoom_Ar}
M. Saini, B. Guthier, H. Kuang, D. Mahapatra, and A.E. Saddik.
\newblock szoom: A framework for automatic zoom into high resolution
  surveillance videos.
\newblock In {\em arXiv preprint arXiv:1909.10164}, 2019.

\bibitem{Schuffler_ABD2013}
P. Sch$\ddot{u}$ffler, D. Mahapatra, J. Tielbeek, F.M. Vos, J. Makanyanga, D.A.
  Pends, C.Y. Nio, J. Stoker, S.A. Taylor, and J.M. Buhmann.
\newblock A model development pipeline for crohns disease severity assessment
  from magnetic resonance images.
\newblock In {\em In Proc: MICCAI-ABD}, 2013.

\bibitem{Schuffler_ABD2014}
P. Sch$\ddot{u}$ffler, D. Mahapatra, J. Tielbeek, F.M. Vos, J. Makanyanga, D.A.
  Pends, C.Y. Nio, J. Stoker, S.A. Taylor, and J.M. Buhmann.
\newblock Semi automatic crohns disease severity assessment on mr imaging.
\newblock In {\em In Proc: MICCAI-ABD}, 2014.

\bibitem{RT36}
Thomas Schlegl, Sebastian~M. Waldstein, Wolf-Dietrich Vogl, Ursula
  Schmidt-Erfurth, and George Langs.
\newblock Predicting semantic descriptions from medical images with
  convolutional neural networks.
\newblock In {\em Proc. Int. Conf. Inform. Process. Med. Imag. (IPMI)}, pages
  437--438, 2015.

\bibitem{Sedai_OMIA18}
S. Sedai, D. Mahapatra, B. Antony, and R. Garnavi.
\newblock Joint segmentation and uncertainty visualization of retinal layers in
  optical coherence tomography images using bayesian deep learning.
\newblock In {\em In Proc. MICCAI-OMIA}, pages 219--227, 2018.

\bibitem{Sedai_MLMI18}
S. Sedai, D. Mahapatra, Z. Ge, R. Chakravorty, and R. Garnavi.
\newblock Deep multiscale convolutional feature learning for weakly supervised
  localization of chest pathologies in x-ray images.
\newblock In {\em In Proc. MICCAI-MLMI}, pages 267--275, 2018.

\bibitem{Sedai_MICCAI17}
S. Sedai, D. Mahapatra, S. Hewavitharanage, S. Maetschke, and R. Garnavi.
\newblock Semi-supervised segmentation of optic cup in retinal fundus images
  using variational autoencoder,.
\newblock In {\em In Proc. MICCAI}, pages 75--82, 2017.

\bibitem{Sedai_EMBC16}
S. Sedai, P.K. Roy, D. Mahapatra, and R. Garnavi.
\newblock Segmentation of optic disc and optic cup in retinal fundus images
  using shape regression.
\newblock In {\em In Proc. EMBC}, pages 3260--3264, 2016.

\bibitem{Sedai_OMIA16}
S. Sedai, P.K. Roy, D. Mahapatra, and R. Garnavi.
\newblock Segmentation of optic disc and optic cup in retinal images using
  coupled shape regression.
\newblock In {\em In Proc. MICCAI-OMIA}, pages 1--8, 2016.

\bibitem{ShinSASHIMI2018}
Hoo-Chang Shin, Neil~A Tenenholtz, Jameson~K Rogers, Christopher~G Schwarz,
  Matthew~L Senjem, Jeffrey~L Gunter, Katherine Andriole, and Mark Michalski.
\newblock {Medical Image Synthesis for Data Augmentation and Anonymization
  using Generative Adversarial Networks}.
\newblock In {\em Proc. MICCAI-SASHIMI}, 2018.

\bibitem{PhS8}
Kihuk Sohn, Honglak Lee, and Xinchen Yan.
\newblock Learning structured output representation using deep conditional
  generative models.
\newblock In {\em Proc. NIPS}, pages 3483--3491, 2015.

\bibitem{DLRevTMI}
Nima Tajbakhsh, Jae.~Y. Shin, Suryakant~R. Gurudu, R.~Todd Hurst,
  Chrostopher~B. Kendall, Michael~B. Gotway, and Jianming Liang.
\newblock Convolutional neural networks for medical image analysis: Full
  training or fine tuning?.
\newblock {\em IEEE Trans. Med. Imag.}, 35(5):1299--1312, 2016.

\bibitem{Tennakoon_OMIA16}
R. Tennakoon, D. Mahapatra, P. Roy, S. Sedai, and R. Garnavi.
\newblock Image quality classification for dr screening using convolutional
  neural networks.
\newblock In {\em In Proc. MICCAI-OMIA}, pages 113--120, 2016.

\bibitem{RT42}
Freerk~G. Venhuizen, Bram van Ginneken, Bart Liefers, Freekje van Asten, Vivian
  Schreur, Sascha Fauser, Carel Hoyng, Thomas Theelen, , and Clara~I. Sanchez.
\newblock Deep learning approach for the detection and quantification of
  intraretinal cystoid fluid in multivendor optical coherence tomography.
\newblock {\em Biomed. Opt. Express}, 9(4):1545--1569, 2018.

\bibitem{VosEMBC}
F.~M. Vos, J. Tielbeek, R. Naziroglu, Z. Li, P. Sch$\ddot{u}$ffler, D.
  Mahapatra, Alexander Wiebel, C. Lavini, J. Buhmann, H. Hege, J. Stoker, and
  L. van Vliet.
\newblock Computational modeling for assessment of {IBD}: to be or not to be?
\newblock In {\em Proc. IEEE EMBC}, pages 3974--3977, 2012.

\bibitem{Xing_MICCAI19}
Y. Xing, Z. Ge, R. Zeng, D. Mahapatra, J. Seah, M. Law, and T. Drummond.
\newblock Adversarial pulmonary pathology translation for pairwise chest x-ray
  data augmentation.
\newblock In {\em In Proc. MICCAI}, pages 757--765, 2019.

\bibitem{RT23}
Xiayu Xu, Kyungmu Lee, Li Zhang, Milan Sonka, and Michael~D. Abramoff.
\newblock Stratified sampling voxel classification for segmentation of
  intraretinal and sub-retinal fluid in longitudinal clinical oct data.
\newblock {\em IEEE Trans. Med. Imag.}, 34(7):1616--1623, 2015.

\bibitem{GAN_MI_Rev}
Xin Yi, Ekta Walia, and Paul Babyn.
\newblock Generative adversarial network in medical imaging: A review.
\newblock {\em Med. Imag. Anal.}, 58, 2019.

\bibitem{Zhao_CVPR2019}
Amy Zhao, Guha Balakrishnan, Fredo Durand, John~V. Guttag, and Adrian~V. Dalca.
\newblock Data augmentation using learned transforms for one-shot medical image
  segmentation.
\newblock In {\em In Proc. CVPR}, pages 8543--8552, 2019.

\bibitem{ZhaoMIA2018}
He Zhao, Huiqi Li, Sebastian Maurer-Stroh, and LiCheng.
\newblock Synthesizing retinal and neuronal images with generative adversarial
  nets.
\newblock {\em Med. Imag. Anal}, 49:14--26, 2018.

\bibitem{Mahapatra_MLMI15_Optic}
J. Zilly, J. Buhmann, and D. Mahapatra.
\newblock Boosting convolutional filters with entropy sampling for optic cup
  and disc image segmentation from fundus images.
\newblock In {\em In Proc. MLMI}, pages 136--143, 2015.

\bibitem{Zilly_CMIG_2016}
J. Zilly, J.M. Buhmann, and D. Mahapatra.
\newblock Glaucoma detection using entropy sampling and ensemble learning for
  automatic optic cup and disc segmentation.
\newblock {\em In Press Computerized Medical Imaging and Graphics},
  55(1):28--41, 2017.

\end{thebibliography}
}

\end{document}